\documentclass[pra,twocolumn,groupedaddress,a4paper,article]{revtex4}

\usepackage{hyperref}
\usepackage{subfigure}
\usepackage{amssymb,amsmath}
\usepackage{graphicx}
\usepackage{color}

\newcommand{\ket}[1]{| #1 \rangle}
\newcommand{\bra}[1]{\langle #1 |}

\newcommand{\fr}[1]{\frac{1}{#1}}
\newcommand{\ketbra}[2]{| #1 \rangle \langle #2 |}
\newcommand{\braopket}[3]{\langle #1 | #2 | #3 \rangle}

\newcommand{\comment}[1]{}
\newcommand{\hidden}[1]{}

\newcommand{\iv}{^{{-}1}}

\newcommand{\akltstate}{\ket{\textsc{AKLT}}}
\newcommand{\groundstate}{\ket{\psi_0}}
\newcommand{\groundstated}{\ket{{\psi}_0(\delta)}}
\newcommand{\akltstated}{\groundstated}
\newcommand{\hamp}[1]{\tilde{H}(#1)}
\newcommand{\hamd}{\hamp{\delta}}
\newcommand{\akltham}{H}
\newcommand{\aklthamd}{\hamd}
\newcommand{\aklthamthree}{H^{(3)}}
\newcommand{\aklthamthreed}{\tilde{H}^{(3)}(\delta)}

\begin{document}
\title{Graph states as ground states of two-body frustration-free Hamiltonians}

\author{Andrew S. Darmawan}
\affiliation{Centre for Engineered Quantum Systems, School of Physics, The University of Sydney, Sydney, NSW 2006, Australia}
\author{Stephen D. Bartlett}
\affiliation{Centre for Engineered Quantum Systems, School of Physics, The University of Sydney, Sydney, NSW 2006, Australia}

\begin{abstract}
The framework of measurement-based quantum computation (MBQC) allows us to view the ground states of local Hamiltonians as potential resources for universal quantum computation. A central goal in this field is to find models with ground states that are universal for MBQC and that are also natural in the sense that they involve only two-body interactions and have a small local Hilbert space dimension.  
Graph states are the original resource states for MBQC, and while it is not possible to obtain graph states as exact ground states of two-body Hamiltonians, here we construct two-body frustration-free Hamiltonians that have arbitrarily good approximations of graph states as unique ground states.  The construction involves taking a two-body frustration-free model that has a ground state convertible to a graph state with stochastic local operations, then deforming the model such that its ground state is close to a graph state. Each graph state qubit resides in a subspace of a higher dimensional particle.
This deformation can be applied to two-body frustration-free Affleck-Kennedy-Lieb-Tasaki (AKLT) models, yielding Hamiltonians that are exactly solvable with exact tensor network expressions for ground states. For the star-lattice AKLT model, the ground state of which is not expected to be a universal resource for MBQC, applying such a deformation appears to enhance the computational power of the ground state, promoting it to a universal resource for MBQC. Transitions in computational power, similar to percolation phase transitions, can be observed when Hamiltonians are deformed in this way.
Improving the fidelity of the ground state comes at the cost of a shrinking gap. While analytically proving gap properties for these types of models is difficult in general, we provide a detailed analysis of the deformation of a spin-1 AKLT state to a linear graph state.

\end{abstract}

\maketitle
\section{Introduction}
In measurement-based quantum computation (MBQC), universal quantum computation is realised by measuring individual particles in a specially entangled multipartite state called a universal resource state. A goal of recent research has been to find local Hamiltonians with ground states that are universal resource states. This could lead to efficient means of preparing resources and, for gapped models, could provide a natural way of protecting resources against thermal errors. 

If such proposals are to eventually be realised in an experiment, it is necessary to put restrictions on the Hamiltonians such that they better resemble physical spin systems. One basic restriction is that the Hamiltonian is two-body, i.e.~each interaction term acts non-trivially on at most two particles. A number of two-body models have been found to have ground states universal for MBQC \cite{bartlett_simple_2006, chen_gapped_2009, cai_universal_2010, wei_affleck-kennedy-lieb-tasaki_2011, miyake_quantum_2011, li_thermal_2011}. 

Furthermore, any specific Hamiltonian can only be approximately realised in an experiment. Therefore, for MBQC it is more useful to look at ground states of families of Hamiltonians, for instance those connected by small perturbations, rather than specific Hamiltonians. A central question is whether universality for MBQC can be regarded as a robust property of a family of Hamiltonians, akin to properties characterising quantum phases of matter. We will call a connected region in the parameter space of a Hamiltonian with ground states universal for MBQC a \emph{computational phase}. Previous work on computational phases has investigated the cluster model in external fields (also at non-zero temperature) \cite{barrett_transitions_2008, doherty_identifying_2009}, with competing Ising interactions \cite{son_quantum_2011}, in three dimensions at non-zero temperature \cite{raussendorf_fault-tolerant_2005}, with error-suppressing interacting cluster terms \cite{fujii_measurement-based_2013}, and under general symmetry preserving perturbations \cite{else_symmetry-protected_2012, else_symmetry_2012-1}. For two-body models, studies have looked at the Haldane phase in one dimension \cite{bartlett_quantum_2010-1}, an anisotropic AKLT model on a honeycomb lattice \cite{darmawan_measurement-based_2012} and versions of the Hamiltonians in Ref. \cite{li_thermal_2011}, which can also be universal at non-zero temperature \cite{fujii_topologically_2012, wei_transitions_2014}. 

Many of the above-mentioned models are frustration-free, meaning that the ground state of the Hamiltonian also minimises the energy of each interaction term. This is useful for MBQC because it means that interaction terms can be switched off without causing the ground state to evolve. Interaction terms must be switched off before a particle is measured or the post-measurement state, no longer an eigenstate, will evolve in time. (We note, however, that always-on interactions may be compensated for in some simple models by measuring at a characteristic clock speed \cite{jennings_quantum_2009, li_thermal_2011}. Furthermore, frustration-freeness is not always needed to adiabatically switch off individual interactions without aversely affecting the ground state \cite{miyake_quantum_2010-1}.) Frustration-free Hamiltonians are also convenient for analytical study because their ground states are often exactly solvable with properties that can be computed easily with tensor network methods \cite{perez-garcia_matrix_2006, perez-garcia_peps_2007}.

Much of the motivation for finding new models for MBQC comes from the fact that graph states, the original universal resource states for MBQC \cite{raussendorf_one-way_2001}, do not appear to arise as ground states of natural Hamiltonians. A graph state $\ket{G}$ of $N$ qubits, defined with respect to a graph $G$ of $N$ vertices, is the state obtained by placing a qubit in the $\ket{{+}}=1/\sqrt{2}(\ket{0}+\ket{1})$ state at each vertex, and applying a controlled-Z gate ${\rm CZ}=\exp{\left(i\pi \ketbra{11}{11}\right)}$ to any pair of qubits connected by an edge. Any graph state is a stabilizer state, and is therefore the unique ground state of a Hamiltonian defined simply as the negative sum of its stabilizer generators. However, these stabilizer Hamiltonians are three-body for one dimensional graph states and at least four-body for two dimensional graph states, and are thus unnatural. 

One might hope for a two-body Hamiltonian with a graph state as a unique ground state, however, for spin-1/2 Hamiltonians, this has been proven impossible \cite{nielsen_cluster-state_2006, van_den_nest_graph_2008}. Frustration-free Hamiltonians are even more restrictive: the ground space of any frustration-free two-body spin-1/2 Hamiltonian is unentangled \cite{bravyi_efficient_2006, chen_no-go_2011}.

One proposal for bypassing these no-go results is using perturbation theory, where ground states are approximate, rather than exact, graph states. The Hamiltonians in these proposals are two-body with spin-1/2 particles however the local Hilbert space dimension is effectively enlarged by either encoding graph state qubits on multiple physical particles \cite{bartlett_simple_2006, griffin_spin_2008}, or by using ancilla particles \cite{van_den_nest_graph_2008}. 
By reducing the perturbation parameter in these models, the ground state can be made arbitrarily close to, but never exactly equal to, the target state at the cost of a shrinking gap. Unlike many of the other schemes mentioned above, these Hamiltonians are necessarily frustrated.

In this paper we present a different type of two-body Hamiltonian that has an approximate graph state as a unique ground state, but, unlike previous perturbative constructions, is frustration-free. 
To obtain such a Hamiltonian, we take a frustration-free Hamiltonian on higher dimensional particles (e.g. spin-3/2) with a ground state convertible to a graph state with stochastic local operations, then deform it such that its ground state is arbitrarily close to, but never exactly equal to, a graph state. Each graph state qubit is encoded into a two dimensional subspace of each physical particle. 

In section \ref{s:2-Ddeformation} we will show how this deformation can drive a spin model into a computational phase, i.e.~a region where the ground states are universal for MBQC. We will illustrate this with an concrete example: the spin-3/2 Affleck-Kennedy-Lieb-Tasaki (AKLT) model on a star lattice \cite{wei_quantum_2013-1}. The ground state of this model is not suspected to be universal for MBQC, unlike other trivalent AKLT states or a graph state on the same lattice. The Hamiltonian deformation applied to this model smoothly transforms the ground state from the star-lattice AKLT state to the star-lattice graph state. We observe a transition in computational power when the ground state becomes sufficiently close to a graph state. Errors arising from not having an exact graph state as a ground state manifest as defects in the resulting graph, which do not affect computational universality if they occur with sufficiently low probability. We will generalise this approach and show how a variety of spin models can be driven into a computational phase. 

Having explored computational aspects of this type of deformation, in section \ref{s:spectral_properties} we will direct our attention to other properties of interest. One crucial property is the spectral gap. For two-body Hamiltonians, due to the no-go results stated above, there must be a trade-off between the fidelity of the ground state with a graph state and the gap. Proving the existence of a gap is, unfortunately, difficult in general. Thus, while our interest is ultimately in two dimensional systems which can be universal for MBQC, in this paper we will restrict most of our discussion of spectral properties to 1-D systems where precise statements can be made. In this section we will also provide some discussion about elementary excitations in these models and the spectral differences between two-body and higher-body Hamiltonians. 

We will conclude with a discussion of future directions for research in section \ref{s:conclusions}.

\section{Model definitions and notation}
In this section we will define families of frustration-free Hamiltonians that have graph states as approximate ground states. First we will describe the importance of local operations and classical communication (LOCC) and quantum state reduction in MBQC. We will then use these ideas to construct frustration-free Hamiltonians with graph states as approximate ground states.

\subsection{Local conversion to graph states}
In the framework of MBQC, aside from the initial preparation of a resource state, the only allowed operations are local operations and classical communication (LOCC). Here we will outline how certain states can be related to graph states by LOCC, an idea that we will use in the next section to construct frustration-free Hamiltonians with graph states as approximate ground states. 

A useful property that all known universal resource states possess is that they can be efficiently converted to universal graph states with single-particle measurements. If one can show that an $N\times N$ cluster state (a graph state on a square lattice) can be obtained by applying LOCC to an entangled state $\ket{\psi}$ of ${\rm poly}(N)$ particles, then we have a proof that $\ket{\psi}$ is a universal resource for MBQC. This conversion, called quantum state reduction \cite{chen_quantum_2010}, can be used to show that the following states are universal resources for MBQC: the tricluster state \cite{chen_gapped_2009}, spin-3/2 AKLT quasi-chains \cite{cai_quantum_2009, wei_quantum_2011}, spin-3/2 AKLT states on a variety of 2-D lattices \cite{wei_affleck-kennedy-lieb-tasaki_2011, wei_quantum_2013-1} spin-3/2 and spin-2 AKLT states with a commuting structure \cite{li_thermal_2011} and certain mixtures of spin-2 and lower spin AKLT states \cite{wei_spin_2013}. The resource states described in \cite{gross_measurement-based_2007}, whose universality was proven using tensor network methods, are universal state preparators \cite{cai_quantum_2009} and thus can also be used to efficiently prepare cluster states.

We can define a more general class of states by those that can be converted to an $N\times N$ cluster state with LOCC but with no restriction on efficiency. Included in this class are AKLT states with higher dimensional spins (e.g. spin-2), the spin-3/2 AKLT state on a star lattice \cite{wei_quantum_2013-1} and the spin-3/2 AKLT state on a honeycomb lattice that has been deformed into a N\'{e}el ordered phase \cite{niggemann_quantum_1997, darmawan_measurement-based_2012}. In each of these examples, no deterministic measurement procedure is known to convert the states efficiently into $N\times N$ cluster states. However, given an exponential amount of time and an exponentially large copy of the state, we can obtain an $N\times N$ cluster state with LOCC. Because of the exponential overhead involved in converting these states to cluster states, we cannot conclude that these states are universal resources for MBQC. 

One final class of states that we mention consists of states that can be converted to 1-D graph states, but not universal graph states, with LOCC. This includes 1-D AKLT states  \cite{affleck_valence_1988} and quantum computational wires \cite{gross_quantum_2010}. To use these states for universal quantum computation, additional operations other than LOCC are required. 

Many of the states mentioned in this section are unique ground states of two-body frustration-free Hamiltonians. This is an appealing property, and is why they are studied as alternative resource states to graph states. In the following, we will show that these two-body Hamiltonians may be deformed such that their ground states are approximate graph states.

\subsection{Frustration-free Hamiltonians with unique ground states close to graph states}
\label{s:deformationdefs}
Here we will describe a class of frustration-free models that have approximate graph states as unique ground states. To exclude trivial cases, here and throughout this paper we will assume that a graph state has $N\ge3$ qubits, and each qubit has degree $d\ge2$. A convenient property of frustration-free Hamiltonians is that each interaction term can be replaced a projection without changing the ground space. Hence, for many purposes, it is often only necessary to consider frustration-free Hamiltonians that are sums of projectors, rather than generic frustration-free Hamiltonians. A Hamiltonian of this form must have a ground state energy of zero. We will use this property in the following, where we show that graph states cannot be exact ground states of two-body, frustration-free Hamiltonians.

\subsubsection{No-go result for for graph states as exact ground states of two-body frustration-free Hamiltonians}
Here we will show that it is not possible to obtain an \emph{exact} qubit graph state as a unique ground state of a two-body frustration-free Hamiltonian. This is already known for qubit Hamiltonians \cite{chen_no-go_2011}, even without the frustration-free property \cite{nielsen_cluster-state_2006, van_den_nest_graph_2008} and we will provide a simplified proof here for the special case of frustration-free, two-body qubit Hamiltonians, which we will then generalise to higher dimensional particles. For qubit Hamiltonians, we can prove this no-go result using properties of stabilizer states. A graph state $\ket{G}$ of $N$ qubits is a stabilizer state with a generating set of stabilizers given by the set of $N$ Pauli operators $X_i \prod_{j\in n_i} Z_j$ for each vertex $i$, where $n_i$ is the set of vertices neighbouring $i$. Any other generating set of stabilizers for $\ket{G}$ can be obtained by taking products of stabilizers from this set. For a stabilizer state, the entanglement entropy of a region of spins $A$ is the minimum of $\fr{2}|S_{AB}|$, where $S_{AB}$ is the set of stabilizer generators that act non-trivially on both $A$ and its complement $B$ and the minimum is taken over all generating sets for the state \cite{fattal_entanglement_2004}.  Now let $A$ be any pair of particles $i$ and $j$. It is not difficult to see that, provided $i$ and $j$ both have degree greater than two, $|S_{AB}|=4$ and thus the entanglement entropy of $A$ is $2$ (this will hold whether $i$ and $j$ are neighbours or not). This implies that the reduced density operator for these two qubits is full-rank and proportional to the identity (i.e., maximally mixed). Hence, any non-zero interaction term $h_{ij}$ (which we assume without loss of generality is a projection operator) applied to particles $i$ and $j$ will have non-zero energy, i.e. $\braopket{G}{h_{ij}}{G}=\mbox{tr}(\rho_{ij} h_{ij})>0$. In other words, we cannot add a two-body interaction term to the Hamiltonian between particles $i$ and $j$ without adding frustration. This holds true for any pair of particles $i$ and $j$, and hence any two-body Hamiltonian with $\ket{G}$ as a ground state must either be frustrated, or zero. This is sufficient to prove that graph states cannot be unique ground states of two-body, frustration-free qubit Hamiltonians. 

We will now generalise this no-go result to the case where each graph state qubit resides in a subspace of a higher dimensional particle. More precisely, we consider a system of $N$ particles each with a local dimension of at least 2, and a non-trivial graph state $\ket{G}$ of $N$ qubits encoded into the particles such that for each particle $i$, there exists a rank-2 projector $P_i$ such that $P_1\otimes P_2 \otimes \dots \otimes P_N\ket{G}=\ket{G}$. In other words, graph-state qubit $i$ resides in the two-dimensional image of $P_i$, which we will call the logical subspace of that particle. Let $H=\sum_{\langle i, j \rangle} h_{ij}$ be some two-body, frustration-free Hamiltonian with $\ket{G}$ as a zero-energy ground state. Using the same stabilizer argument above, the reduced density operator of any pair of particles $i$ and $j$ is $\rho_{ij}=P_i\otimes P_j$, i.e. it is maximally mixed on the logical subspace of particles $i$ and $j$. The kernel of any interaction term $h_{ij}$ acting between $i$ and $j$ must contain the image of $P_i\otimes P_j$, as otherwise the energy of the interaction term $\braopket{G}{h_{ij}}{G}=\mbox{tr}(h_{ij} (P_i\otimes P_j))$ will be greater than zero (and thus frustrated). However, if the kernel of $h_{ij}$ contains the image of $P_i\otimes P_j$, then any two-qubit state in the logical subspace of particles $i$ and $j$ will have zero energy. As this holds for any pair of particles $i$ and $j$, a two-body Hamiltonian of the form $H=\sum_{\langle i, j \rangle} h_{ij}$, satisfying $h_{ij}\ket{\psi}=0$ for all interaction terms $h_{ij}$ will have at least a $2^N$-fold ground space degeneracy corresponding to the image of $P_1\otimes P_2 \otimes \dots P_N$. Therefore, we cannot obtain an exact graph state as a unique ground state of a two-body, frustration-free Hamiltonian, even when each graph state qubit is encoded into a higher dimensional physical particle. 

\subsubsection{Approximate graph states as ground states}

While having exact graph states as unique ground states of two-body frustration-free Hamiltonians is not possible, it is possible to obtain arbitrarily good approximations to them. Let $H=\sum h_\alpha$ be a local Hamiltonian acting on $N$ particles with positive interaction terms $h_\alpha$ and a unique ground state $\groundstate$. We will assume that the model is frustration-free, with ~$h_\alpha\groundstate=0$ for all $\alpha$. We also assume that there exists a set of rank-2 projection operators $(P_j)_{j=1}^N$ such that $\ket{G}:=P_1\otimes P_2 \otimes \dots \otimes P_N \groundstate$ is an $N$ qubit graph state, where graph state qubit $j$ resides in the two-dimensional image of $P_j$. This condition is satisfied by many of the states mentioned in the previous section (the only exceptions possibly being the universal state preparators which generate graph states in a way that does not preserve locality properties of the original state).

Given such a Hamiltonian and a set of rank-2 projectors $\left(P_j\right)_{j=1}^N$ we define a one-parameter family of operators at each site $j$, which we call deformation operators, by $D_j(\delta)=\delta P_j + (1-P_j)$ where $\delta\ge0$. We will use the fact that $D_j(\delta)$ has an inverse given by $D_j(\delta)\iv=\delta\iv P_j+(I-P_j)$ for $\delta>0$. In order to keep our deformed Hamiltonian in the form as a sum of projectors (for convenience) we define a local map $\mathcal{Q}$ as follows. If $A$ is an operator we define $\mathcal{Q}(A)$ to be the projection onto the orthogonal complement of the kernel of $A$, such that $\mathcal{Q}(A)$ has the same kernel as $A$ but only has eigenvalues 0 and 1. We define the deformed interaction term by
\begin{align}
    \tilde{h}_\alpha(\delta):=\mathcal{Q}\left(\left[\bigotimes_{j\in r_\alpha} D_j(\delta)\right]h_\alpha\left[\bigotimes_{j\in r_\alpha} D_j(\delta)\right]\right)\,,
    \label{e:deformed_interaction}
\end{align}
where $r_\alpha$ is the set of particles on which $h_\alpha$ acts non-trivially. The model $\hamd:=\sum_\alpha \tilde{h}_\alpha(\delta)$ is frustration-free and for $\delta>0$ has a unique ground state given by 
\begin{equation}
    \groundstated:=\left[\bigotimes_{j=1}^N D_j(\delta)^{-1}\right] \groundstate\,.
    \label{e:invertible_gs}
\end{equation}
This state satisfies $\ket{\psi_0(1)}=\groundstate$ and $\lim_{\delta\rightarrow 0}\groundstated=\ket{G}$. Thus, the unique ground state of $\hamd$ can be made arbitrarily close to a graph state $\ket{G}$ by setting $\delta$ sufficiently small. Hence we can define frustration-free Hamiltonians that have graph states as approximate ground states.

While reducing $\delta$ towards 0 increases the fidelity of the ground state with a graph state to 1, the Hamiltonian $\hamp{0}$ obtained by setting $\delta=0$ does not have $\ket{G}$ as a unique ground state. The interaction term $\tilde{h}_\alpha(\delta)$ loses rank at $\delta=0$ (compared to $\delta>0$) and the overall ground space of $\hamp{0}$ is highly degenerate. Hence, as expected, this construction does not allow us to get exact graph states as unique ground states, only approximate ones. 

This construction transforms the ground state in a simple way, especially if the undeformed ground state $\groundstate$ is a projected entangled pair state (PEPS). A PEPS is defined with respect to a graph $G$, where each vertex is associated with a physical particle, and each edge associated with a bond. For each edge $e=(u,v)$ in $G$ place a maximally entangled bond state $\ket{\omega}=\sum_{i=1}^D \ket{ii}$, where $D$ is called the bond dimension, and associate one particle of $\ket{\omega}$ with $u$ and the other with $v$. For each vertex $v$ define an operator $\mathcal{P}_v:(\mathbb{C}^D)^{\otimes p} \rightarrow \mathbb{C}^{d_v}$ that maps all $p$ bond particles associated with vertex $v$ to a $d_v$-dimensional physical particle. The PEPS $\ket{\psi}=\bigotimes_v \mathcal{P}_v \bigotimes_e \ket{\omega}_e$ is then obtained by mapping all bond particles to physical particles. If the undeformed ground state is a PEPS state with operators $\mathcal{P}_v$ then the deformed ground state $\groundstated$ is also a PEPS with operators given simply by $D_v(\delta)\iv \mathcal{P}_v$. We remark that this deformation is similar to the deformation used in Ref. \cite{schuch_classifying_2011} that takes a PEPS to its standard isometric form. Clearly, the deformation cannot change the bond dimension of the PEPS.

The deformation also preserves frustration-freeness and the locality of interaction terms: deforming $k$-local Hamiltonians yields $k$-local Hamiltonians. In particular, two-body Hamiltonians remain two-body Hamiltonians under this deformation. Finally, if the undeformed $\groundstate$ can be converted to a graph state $\ket{G}$ with LOCC, then so can any deformed state $\groundstated$. However, the statistics of this conversion will, in general, vary with $\delta$.

The path of Hamiltonians $\hamd$ specifies a continuous path of ground states $\groundstated$ for $\delta\in(0,1]$. States along this path are related by stochastic local operations and classical communication (SLOCC) \cite{dur_three_2000}. Two $N$-particle states $\ket{\phi}$ and $\ket{\theta}$ are said to SLOCC equivalent if there exists a collection of $N$ invertible local operators $(A_j)_{j=1}^N$ such that $A_1\otimes A_2 \otimes \dots \otimes A_N \ket{\phi}=\ket{\theta}$. The operational meaning of two states $\ket{\phi}$ and $\ket{\theta}$ being SLOCC equivalent is that it is possible to convert $\ket{\phi}$ to $\ket{\theta}$ and vice-versa with LOCC with a non-vanishing (although possibly exponentially small) probability of success. From Eq. \eqref{e:deformed_interaction}, we see that varying $\delta>0$ keeps the ground state within the SLOCC class of the undeformed ground state $\groundstate$. By replacing the deformation operators $\left(D_j(\delta)\right)_{j=1}^N$ by arbitrary positive invertible operators $\left(A_j\right)_{j=1}^N$ in the definition of $\tilde{h}_\alpha(\delta)$ we can construct a frustration-free parent Hamiltonian for any state within the SLOCC class of $\groundstate$.  

A previous study has investigated the computational universality of SLOCC deformed graph states \cite{dsouza_strategies_2011}. In this work, it was shown that under particular types of deformations, graph states (on certain graphs) will remain universal for MBQC. In contrast, the SLOCC classes we consider contain arbitrarily good approximations to graph states, but not exact graph states. It should also be noted that SLOCC equivalence of two states does not imply similar physical properties. For instance, it has been shown that moving around an SLOCC class can result in a phase transition (indicated by two-point correlation functions) \cite{niggemann_quantum_1997, perez-garcia_peps_2007}.

In the remainder of this paper we will explore how properties of the deformed model $\tilde{H}(\delta)$ change as the ground state is deformed towards a graph state i.e.~as $\delta\rightarrow0$. In Section \ref{s:2-Ddeformation}, we study a 2-D spin system for which reducing $\delta$ to a constant value improves the computational properties of the ground state for measurement-based quantum computation. In Section \ref{s:spectral_properties} we explore how physical properties, such as the energy gap, vary as a function of $\delta$. 

\section{Universality of ground states for MBQC}
\label{s:2-Ddeformation}
Here we will illustrate how a certain class of spin models can be driven into a computational phase characterised by ground states that are universal for MBQC. We will consider the spin-3/2 AKLT model on a star lattice as an example. This model has a ground state that is not expected to be universal for MBQC. Using the deformation described in section \ref{s:deformationdefs} we will show how to drive this model into a phase characterised by ground states that are universal for MBQC. After this example, we will explain how this may be done in general to a variety of frustration-free spin models.

\subsection{The spin-3/2 AKLT model}
\label{s:spin32_aklt}
Here we will briefly review the spin-3/2 AKLT model and how its ground state may converted to a graph state with single-particle measurements, following \cite{wei_affleck-kennedy-lieb-tasaki_2011}. The spin-3/2 AKLT model is a frustration-free model that may be defined on an arbitrary trivalent graph \cite{affleck_valence_1988}. It has a Hamiltonian given by 
\begin{equation}
    \akltham=\sum_{\langle i,j \rangle} P^{J=3}_{ij}\,,
\end{equation}
where $P^{J=3}_{ij}$ is the projector of particles $i$ and $j$ onto the total spin $J=3$ subspace of two spin-3/2 particles, and the summation is carried out over all neighbouring pairs $\langle i,j\rangle$. This model shares some similarities with the Heisenberg antiferromagnet: it has full $SO(3)$ rotational symmetry and it penalises alignment of neighbouring spins. However, the AKLT model and the Heisenberg model have some significant differences. For instance, on a honeycomb lattice the AKLT model has has exponentially decaying correlation functions and there is strong numerical evidence of a gap (although this has not been analytically proven) \cite{garcia-saez_spectral_2013-1} while the Heisenberg model on the same lattice has N\'{e}el order and is therefore gapless. One reason why the AKLT model is easier to study than the Heisenberg model is that it has an exactly solvable ground state with a simple PEPS description which we hereafter call the AKLT state $\akltstate$.

\begin{figure}
\centering
\mbox{(a)\subfigure{\includegraphics[width=0.14\textwidth]{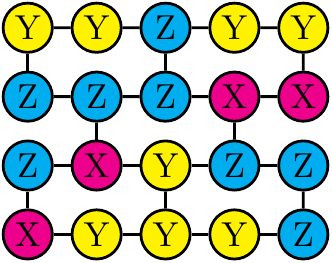}}(b)\subfigure{\includegraphics[width=0.14\textwidth]{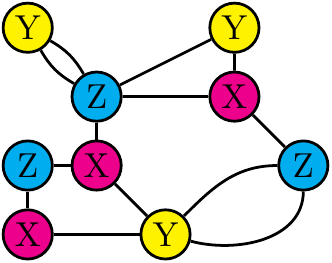}}(c)\subfigure{\includegraphics[width=0.14\textwidth]{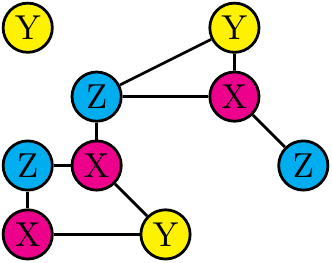}}}
\caption{Illustration of how the graph state is encoded on the post-reduction AKLT state. In (a) we have a small 2-D AKLT state on a trivalent lattice, where each node corresponds to a spin-3/2 particle, and is labelled according to the reduction outcome obtained. In (b) nodes now represent domains of like outcomes, and edges are bonds between domains. In (c) we have illustrated the encoded graph state, obtained from (b) by deleting edges modulo 2. Each node represents an encoded qubit, and each edge a graph state edge.}
\label{f:outcomestographstate}
\end{figure}

The AKLT state can be converted to an encoded graph state with single-particle measurements \cite{wei_affleck-kennedy-lieb-tasaki_2011, miyake_quantum_2011}. This is done by applying a three outcome measurement to each spin-3/2 particle. The measurement operators are defined as follows. Let $\ket{m}_b$ be the spin-3/2 state satisfying $S_b\ket{m}_b=m\ket{m}_b$ where $S_b$ is the spin-3/2 component along the $b$ axis, $b\in\{x,y,z\}$, and $m\in\{\frac{3}{2}, \frac{1}{2}, {-}\frac{1}{2}, {-}\frac{3}{2}\}$. The measurement operators for the initial reduction are chosen to be $\{F^x, F^y, F^z\} $ where, for $b\in \{x,y,z\}$,
\begin{align}
F^b=\sqrt{\frac{2}{3}}\left(\ket{\tfrac{3}{2}}_b\bra{\tfrac{3}{2}}+\ket{{-}\tfrac{3}{2}}_b\bra{{-}\tfrac{3}{2}}\right)\,.
\label{e:undeformed_measurement}
\end{align}
These operators satisfy the relation $\sum_{b=x,y,z}F^{b\dag} F^b=I $ and therefore form a valid POVM. Each particle is measured, and a graph state is encoded on the post-reduction state as follows (we have illustrated the encoding in Fig.\ \ref{f:outcomestographstate}). A \emph{domain} is defined as a connected set of particles with the same measurement outcome.  Each domain encodes a single qubit in the graph state. An edge exists between two encoded qubits if an odd number of bonds (in the original lattice) connect the corresponding domains. 

\begin{figure}
\centering
\includegraphics[width=0.4\textwidth]{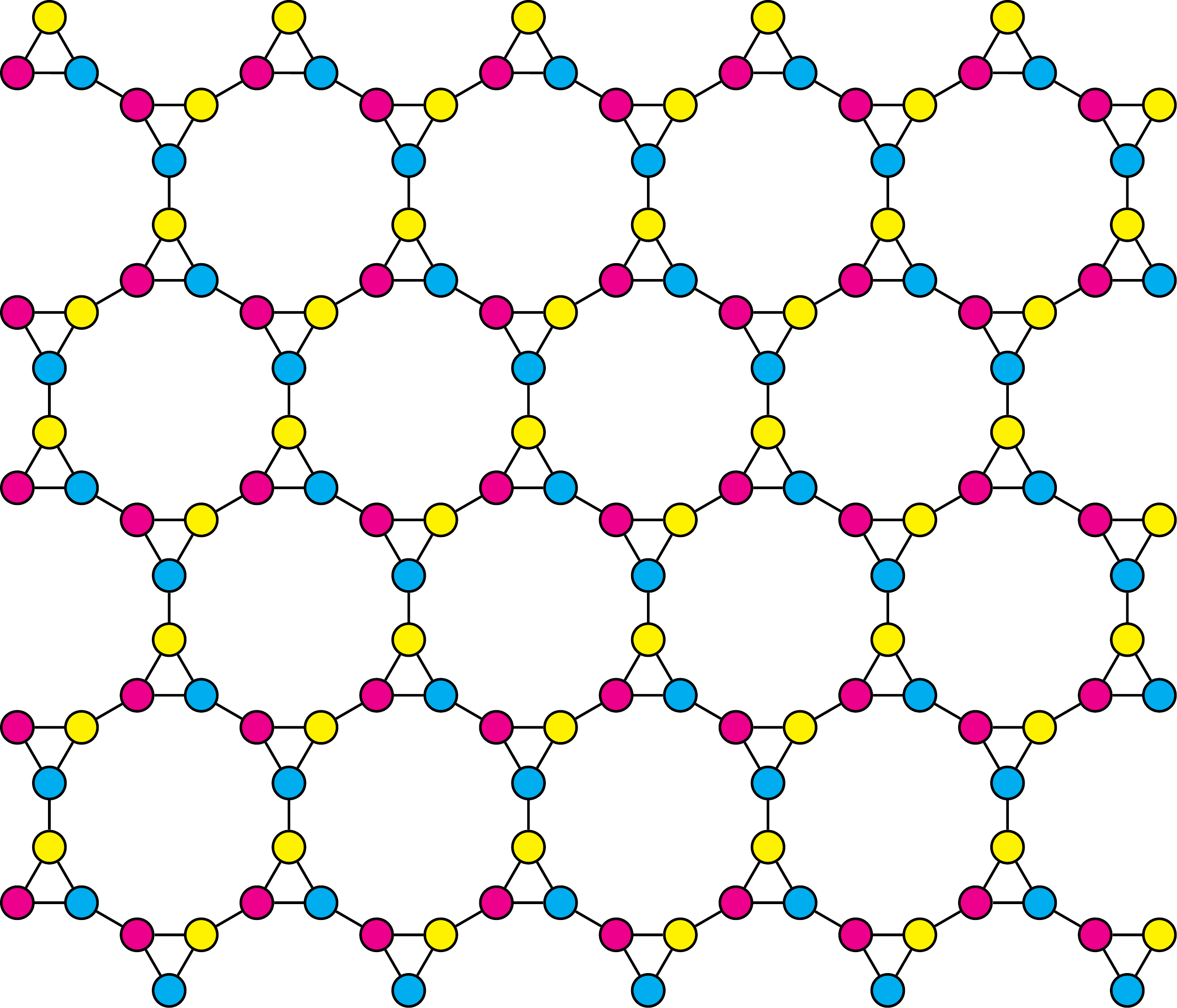}
\caption{The $(3,12^2)$ lattice, which we call the star lattice, is an Archimedian tiling of the plane characterised by one triangle and two dodecagons surrounding each vertex. We have illustrated a particular three-colouring of the lattice.}
\label{f:star-lattice-colouring}
\end{figure}

The resulting graph state will have a graph structure determined by the measurement outcomes and the graph or lattice on which the AKLT model was defined. We will specify lattices using their vertex configuration so, for instance, a lattice specified by $(4,8^2)$ has a square and two octagons surrounding each vertex. It has been shown that ground states of the $(6^3)$ honeycomb,  $(4,8^2)$ square-octagon and $(4,6,12)$ cross-lattice AKLT models efficiently yield cluster states that are universal for MBQC \cite{wei_affleck-kennedy-lieb-tasaki_2011, wei_quantum_2013-1}. However, the ground state of the $(3,12^2)$ star-lattice AKLT model (shown in Fig. \ref{f:star-lattice-colouring}) yields graph states that do not possess the right connectivity properties for universal MBQC (they are not reducible to square lattice cluster states by single-particle measurement). This is in contrast to a graph state defined on the star lattice, which is universal for MBQC due to the fact that it can be converted to a cluster state with single-particle measurements. Thus, the star-lattice AKLT state appears to have different  computational properties to the star-lattice graph state. We will use this example to show how we can enhance the computational properties of the star-lattice AKLT state by applying the deformation described in Sec. \ref{s:deformationdefs}

\subsubsection{Deforming the spin-3/2 AKLT model}
We will now deform the star-lattice spin-3/2 AKLT model such that its ground state is close to the star lattice graph state $\ket{G}$. Each particle is assigned one of three deformation axes $\{x,y,z\}$, such that no two neighbouring particles have the same deformation axis. In other words, the deformation axes colour the lattice. We will use the colouring illustrated in Fig. \ref{f:star-lattice-colouring}, however any three-colouring of the star lattice will do. Let $c_j\in\{x,y,z\}$ be the deformation axis of particle $j$. From the reduction procedure described in Sec. \ref{s:spin32_aklt}, we observe that the set of projectors 
\begin{equation}
P_j=\ket{\tfrac{3}{2}}_{c_j}\bra{\tfrac{3}{2}}+\ket{{-}\tfrac{3}{2}}_{c_j}\bra{{-}\tfrac{3}{2}}\,,
\end{equation} 
defined for all sites $j=1,\dots,N$ will convert the star-lattice AKLT state to a star-lattice graph state in the sense that $\bigotimes_{j=1}^N P_j \akltstate=\ket{G}$ (where graph state qubit $j$ resides in the logical subspace of particle $j$). As $\{c_j\}$ colours the lattice, each qubit of this graph state resides in a 2-D subspace of a single spin-3/2 particle. 

Using this set of projectors, we can follow the construction described in Sec. \ref{s:deformationdefs} to deform the Hamiltonian such that its ground state is arbitrarily close the star-lattice graph state. We define a set of deformation operators by $D_j(\delta)=\delta P_j + (1-P_j)$ for $j=1, \dots, N$. The deformed model is then given by 
\begin{align}
    \aklthamd=\sum_{\langle i,j \rangle}\mathcal{Q}\left(\left[D_i(\delta) \otimes D_j(\delta)\right]P_{ij}^{J=3}\left[D_i(\delta) \otimes D_j(\delta)\right]\right)\,,
    \notag
\end{align}
where $\mathcal{Q}$ is defined such that $\mathcal{Q}(A)$ is the projection onto the orthogonal complement of the kernel of $A$. 
The model $\aklthamd$ has a ground state that can be written down exactly as
\begin{equation}
    \akltstated:=\left[\bigotimes_{j=1}^N D_j(\delta)^{-1}\right] \akltstate\,.
\end{equation}
This ground state satisfies $\lim_{\delta\rightarrow0}\akltstated=\ket{G}$, i.e.~we can make the ground state arbitrarily close to a graph state by setting $\delta$ to be sufficiently small.  

However, it is not necessary to reduce $\delta$ significantly for the purpose of MBQC. We will show, in later sections, that it is possible to make a ground state universal for MBQC by reducing $\delta$ to an $N$-independent constant, which need not be very small (in the example considered, we find a transition at $\delta\approx0.5$). 

To summarise, in this section we showed that the star-lattice AKLT model can be deformed such that its ground state approaches a graph state. In the following section we  will study how the statistics of state reduction on $\akltstated$ change as $\delta$ is reduced.

\subsubsection{Reducing deformed ground states to graph states}
\label{s:reducing_groundstates}
Here we will show how the reduction measurement, defined in Eq. \eqref{e:undeformed_measurement}, can be altered to work for the deformed AKLT states defined in the previous section. Specifically, we will define a reduction procedure that transforms the ground state $\akltstated$ for any $\delta\in(0,1]$ to a graph state with local operations. By looking at the statistics of this reduction, we will find that the graph states obtained for different values of $\delta$ have different computational power. This will allow us to identify a computational phase containing ground states that are universal for MBQC. 

The reduction procedure involves applying a site-dependent and $\delta$-dependent measurement to each particle. We measure particle $j$ with a measurement defined by three operators $\{\tilde{F}^{x}_j(\delta),\tilde{F}^{y}_j(\delta),\tilde{F}^{z}_j(\delta)\}$ which we define for $b\in\{x,y,z\}$ as
\begin{equation}
    \tilde{F}^{b}_j(\delta):=\left(\frac{3-\delta^2}{2\delta^2}\right)^{\frac{\delta_{b,c_j}}{2}}F^b D_j(\delta)\,,
    \label{e:deformed_measurement}
\end{equation}
where $\delta_{b,c_j}$ is 1 if $b=c_j$ and 0 otherwise, $\{F^x,F^y,F^z\}$ are defined in Eq. \eqref{e:undeformed_measurement} and $D_j(\delta)$ is the site dependent deformation operator defined in the previous section. One can check that these operators form a valid POVM i.e.~$\tilde{F}^{x}_j(\delta)^\dag\tilde{F}^{x}_j(\delta)+\tilde{F}^{y}_j(\delta)^\dag\tilde{F}^{y}_j(\delta)+\tilde{F}^{z}_j(\delta)^\dag\tilde{F}^{z}_j(\delta)=I$ for any $\delta\in(0,1]$ and any $j$. This is similar to the measurement defined in \cite{darmawan_measurement-based_2012} to convert deformed AKLT states to graph states. 

The collection of measurement outcomes, after measuring all particles, can be represented as a string $\sigma=\sigma_1\sigma_2\dots\sigma_N$, where $\sigma_j\in\{x,y,z\}$ labels which of the three outcomes $\{\tilde{F}^{x}_j(\delta),\tilde{F}^{y}_j(\delta),\tilde{F}^{z}_j(\delta)\}$ was obtained at site $j$. The transformed ground state after measurement will be 
\begin{align}
    &\left[\bigotimes_{j=1}^N\tilde{F}^{\sigma_j}_j(\delta)\right] \akltstated\,,\notag\\
    &=\left(\frac{3-\delta^2}{2\delta^2}\right)^{\fr{2}N_m(\sigma)} \left[\bigotimes_{j=1}^N F^{\sigma_j}D_j(\delta) D_j(\delta)^{-1}\right]\akltstate\,,\notag\\
    &=\left(\frac{3-\delta^2}{2\delta^2}\right)^{\fr{2}N_m(\sigma)}\left[\bigotimes_{j=1}^N F^{\sigma_j}\right]\akltstate\label{e:post_measurement}\,.
\end{align}  
where $N_m(\sigma)=\sum_{j=1}^N\delta_{c_j,\sigma_j}$ equals the number of sites for which the deformation axis $c_j$ matches the outcome $\sigma_j$ and we note that we have left the state unnormalised. Thus the post-measurement state for a given outcome $\sigma$ is independent of $\delta$ (ignoring its amplitude), and is identical to the state obtained from the undeformed model at $\delta=1$. This state encodes a graph state on a graph that is determined by the outcomes $\sigma$. However, as the relative amplitudes of these states are $\delta$-dependent, the probability of getting any particular $\sigma$ is also $\delta$-dependent. We will investigate this probability distribution in the following section. 

\subsubsection{Statistics of reduction procedure}
Here we will describe the statistics of the reduction procedure described above. It was shown by Wei \textit{et al}. \cite{wei_affleck-kennedy-lieb-tasaki_2011, wei_quantum_2013-1} that the probability $p_1(\sigma)$ of obtaining a reduction outcome $\sigma$ for the undeformed, star-lattice AKLT state is given by 
\begin{equation}
    p_{\delta=1}(\sigma)=\fr{\mathcal{Z}}h(\sigma)2^{|V(\sigma)|-|E(\sigma)|}\,,
    \label{e:distribution}
\end{equation}
where $h(\sigma)$ equals 0 when $\sigma$ contains a domain that is not two-colourable and equals 1 otherwise, $|V(\sigma)|$ is the number of domains in $\sigma$, $|E(\sigma)|$ is the number of inter-domain edges in $\sigma$ and $\mathcal{Z}=\sum_\sigma h(\sigma)2^{|V(\sigma)|-|E(\sigma)|}$ is a normalisation factor. The $h(\sigma)$ term arises due to the fact that any domain that is not two-colourable must contain a pair of aligned neighbouring spins $\ket{\tfrac{3}{2}}_b\ket{\tfrac{3}{2}}_b$ or $\ket{{-}\tfrac{3}{2}}_b\ket{{-}\tfrac{3}{2}}_b$, yet the AKLT state is orthogonal to any such state. The presence of $h(\sigma)$ forces domains to be string-like. Using this fact, we can simplify the quantity $L(\sigma):=|V(\sigma)|-|E(\sigma)|$ that appears in Eq. \eqref{e:distribution}. If we imagine flipping a spin in $\sigma$, we see that $L(\sigma)$ will increase by 1 if and only if a loop of like outcomes is created, and will decrease by 1 if and only if a loop of like outcomes is lost. Hence $L(\sigma)$ is equal to the number closed of loops of like outcomes, up to an ignorable additive constant. 

On the star lattice we can show that the $2^{L(\sigma)}$ term has a negligible effect. An intuitive reason for this is as follows. The smallest possible loop contributing to $L(\sigma)$ consists of twelve particles surrounding a dodecagon. Only three of $3^{12}$ possible outcomes on particles surrounding a dodecagon correspond to loops of like outcomes (one outcome for an $x,y$ and $z$ loop separately). The $2^{L(\sigma)}$ term will double the relative probability of such loop outcomes, however the total probability that all outcomes surrounding a given dodecagon match is still only $\approx1.1\times 10^{-5}$. Larger loops will be exponentially suppressed. Hence, for this particular lattice, we can safely ignore the $2^{L(\sigma)}$ term, as it only affects a tiny fraction of all possible measurement outcomes.

If we now vary $\delta$, the probability distribution changes in a simple way. We see in Eq. \eqref{e:post_measurement}, that the relative amplitude of a given outcome $\sigma$ has a factor of $\left((3-\delta^2)/(2\delta^2)\right)^{N_m(\sigma)/2}$ compared with the amplitude of the undeformed post-measurement state. Hence the probability that $\sigma$ is obtained when measuring the ground state of $\aklthamd$ is
\begin{equation}
    p_\delta(\sigma)=\fr{\mathcal{Z}(\delta)}h(\sigma)2^{L(\sigma)}\left((3-\delta^2)/2\delta^2\right)^{N_m(\sigma)}\,,
    \label{e:distribution_deformed}
\end{equation}
where $N_m(\sigma)$, $L(\sigma)$ and $h(\sigma)$ are as above, and $\mathcal{Z}(\delta)=\sum_\sigma h(\sigma)2^{L(\sigma)}\left((3-\delta^2)/2\delta^2\right)^{N_m(\sigma)}$ is a normalisation factor. 

The term $\left((3-\delta^2)/2\delta^2\right)^{N_m(\sigma)}$ is the only term affected by varying $\delta$, aside from the normalisation factor. One can think of this term as a classical external field that energetically favours one of $x,y$ or $z$ at each site, depending on the site's deformation axis. At $\delta=1$, this term is 1 for all $\sigma_j$ i.e., no site is biased towards any particular outcome. In the limit as $\delta\rightarrow0$, the sites will be fully biased towards their deformation axes, i.e.~we will obtain $\sigma_j=c_j\,, \forall j$ with probability 1. 
In this limit, the post-measurement state is $\ket{G}$, the star-lattice graph state, which is known to be universal for MBQC. 
We cannot, however, set $\delta=0$ as the ground space becomes degenerate at this point.
For any fixed, non-zero $\delta$, the probability of obtaining $\sigma_j=c_j$ (corresponding to an exact star-lattice graph state) decays exponentially in the system size. Fortunately, it is not necessary for the reduction to produce an exact star-lattice graph state, as graph states on regular lattices can have a non-zero density of defects and still be universal for MBQC \cite{browne_phase_2008}. As we will explore in the following section, for any fixed $\delta$ the density of defects is roughly constant in the system size, and when $\delta$ is below some critical, system-size-independent value, the ground state will be a universal resource for MBQC.

To summarise, in this section we outlined the reduction procedure that converts the ground state of $\aklthamd$ into a graph state, and we explained how the statistics of this process depend on the deformation parameter $\delta$. In the following section we will investigate the effect that this has on the computational properties of the groundstate for MBQC. 

\subsubsection{Finding a computational phase transition}
Next we will investigate the computational universality of the ground state of $\aklthamd$ as $\delta$ is varied. Recall that, as $\delta$ varies in the range $(0,1]$, the ground state of $\aklthamd$ varies smoothly from the star-lattice AKLT state (which is not expected to be universal for MBQC) to the star-lattice graph state (which is universal for MBQC). Here we will show that there is a transition in computational power at an intermediate value of $\delta$. 

Recall that we use the ground states of $\aklthamd$ for MBQC by first reducing them to graph states, a process which is stochastic with probabilities determined by Eq. \eqref{e:distribution_deformed}. Two conditions, stated in \cite{wei_affleck-kennedy-lieb-tasaki_2011}, will ensure that that such a reduction procedure can efficiently produce cluster states that are universal for MBQC:
\begin{description}
    \item[C1]The size of the largest domain scales at most logarithmically with the system size.
    \item[C2]The probability that the resulting graph state has a crossing path tends to 1 in the thermodynamic limit.
\end{description}
The first condition C1 will ensure that the number of qubits in the resulting graph state is extensive. The second condition C2, combined with C1, will ensure the existence of a macroscopic number of distinct crossing paths in both lattice dimensions. In short, C1 ensures that the resulting graph state has sufficiently many qubits, while C2 ensures that the resulting graph states have sufficient connectivity for universal MBQC. Together, these conditions ensure that resulting graph states can be efficiently converted to cluster states with single-particle measurements. 

\begin{figure*}
\centering
\mbox{\subfigure{\includegraphics[width=0.45\textwidth]{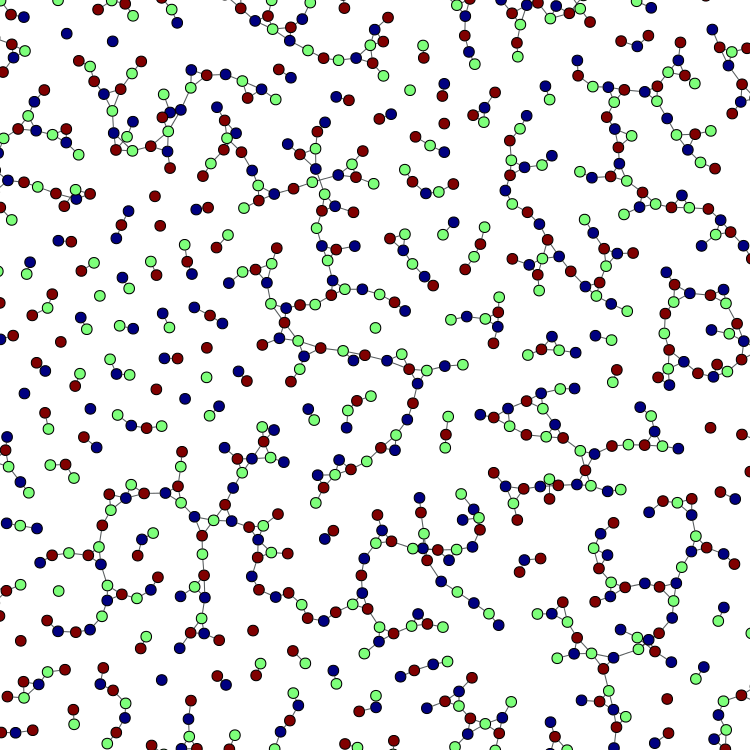}}\quad\quad\subfigure{\includegraphics[width=0.45\textwidth]{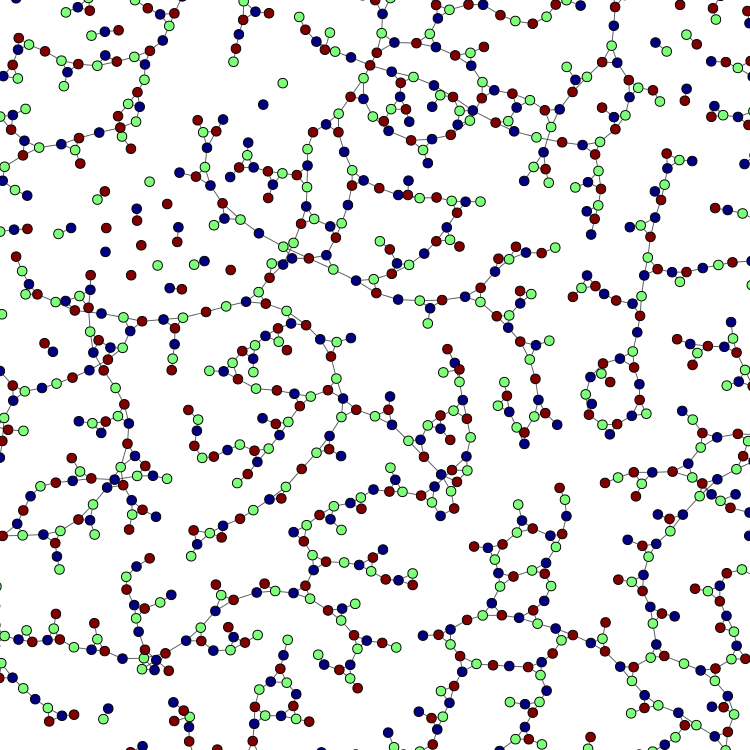}}}
\mbox{\subfigure{\includegraphics[width=0.45\textwidth]{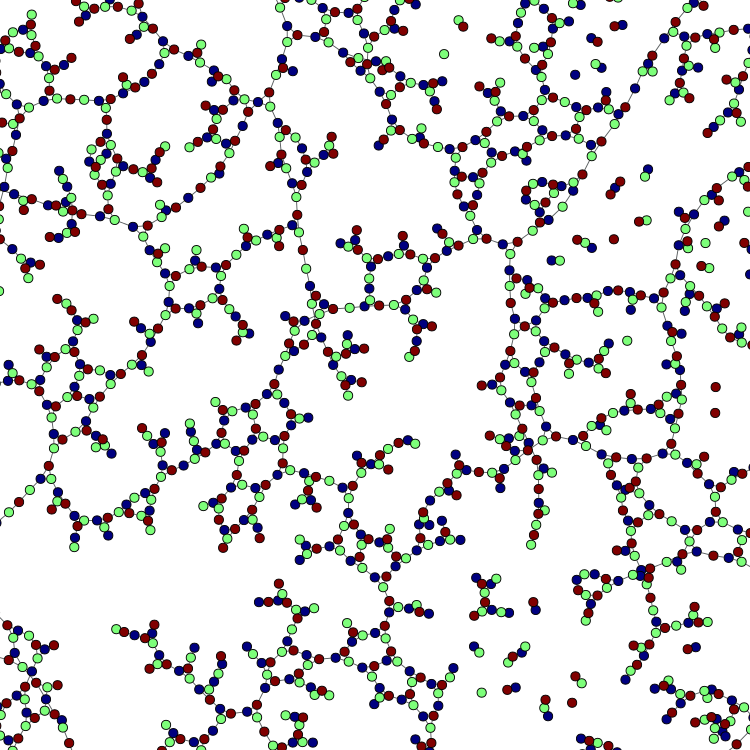}}\quad\quad\subfigure{\includegraphics[width=0.45\textwidth]{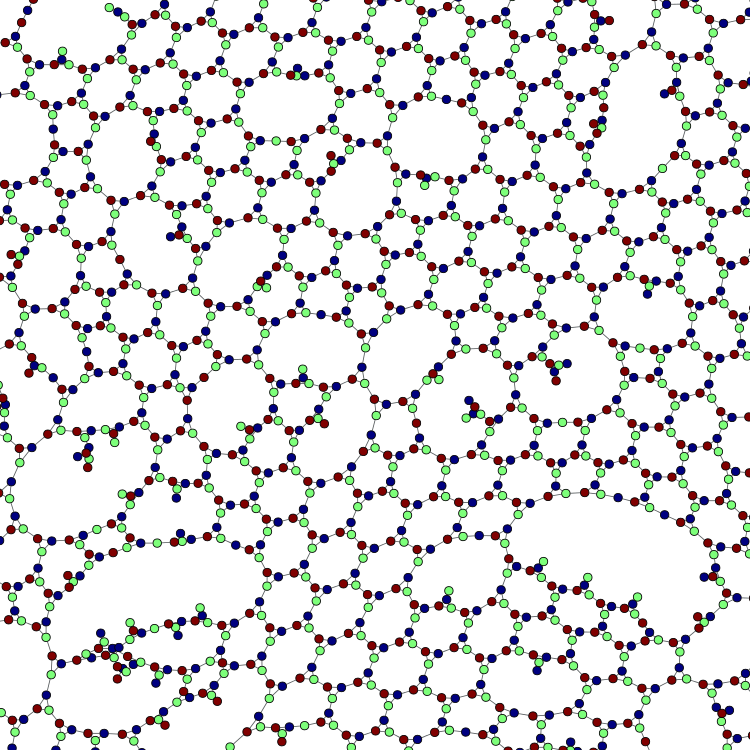}}}
\caption{Sample graphs obtained from reducing ground states of $\aklthamd$ for various values of $\delta$. Note that the vertices pictured correspond to encoded graph state qubits, not physical particles, and we have used a scalable multilevel force directed placement (SFDP) algorithm to compute an aesthetically appealing layout of each graph. The figure on the top left is at $\delta=1$, i.e.~the AKLT point. This is in the non-universal phase where the probability of a macroscopic graph state tends to zero in the thermodynamic limit. The top right figure is at $\delta=0.5$, very close to the transition point. Here we see the first appearance of a macroscopic graph state with a path of connected qubits crossing the entire lattice. At $\delta=0.4$, shown in the bottom left, we are further into the universal phase, with a higher density of crossing paths. Finally, at $\delta=0.2$, shown in the bottom right, we are well into the universal phase. By this point the underlying lattice (the star lattice) has become visible. One way of interpreting the ground states of $\aklthamd$ within the computational phase, is as defective (although still universal) star-lattice graph states, which can most clearly be seen in the $\delta=0.2$ sample.}
\label{f:graphpictures}
\end{figure*}

We calculated whether the ground state of $\hamd$ satisfied these conditions for various values of $\delta\in(0,1]$. This involved sampling graphs from of the $\delta$-dependent distribution Eq. \eqref{e:distribution_deformed} using Monte-Carlo methods, and checking the domain size as well as the existence of a crossing path in both lattice dimensions. We have illustrated samples of post-measurement encoded graph states for different values of $\delta$ in Fig. \ref{f:graphpictures}.

We found that C1 is easily satisfied for all values of $\delta\in(0,1]$. Our Monte-Carlo sampling showed that the domain sizes remain small, and in fact shrink to 1 as $\delta$ tends to zero. 

\begin{figure}
    \includegraphics[width=0.5\textwidth]{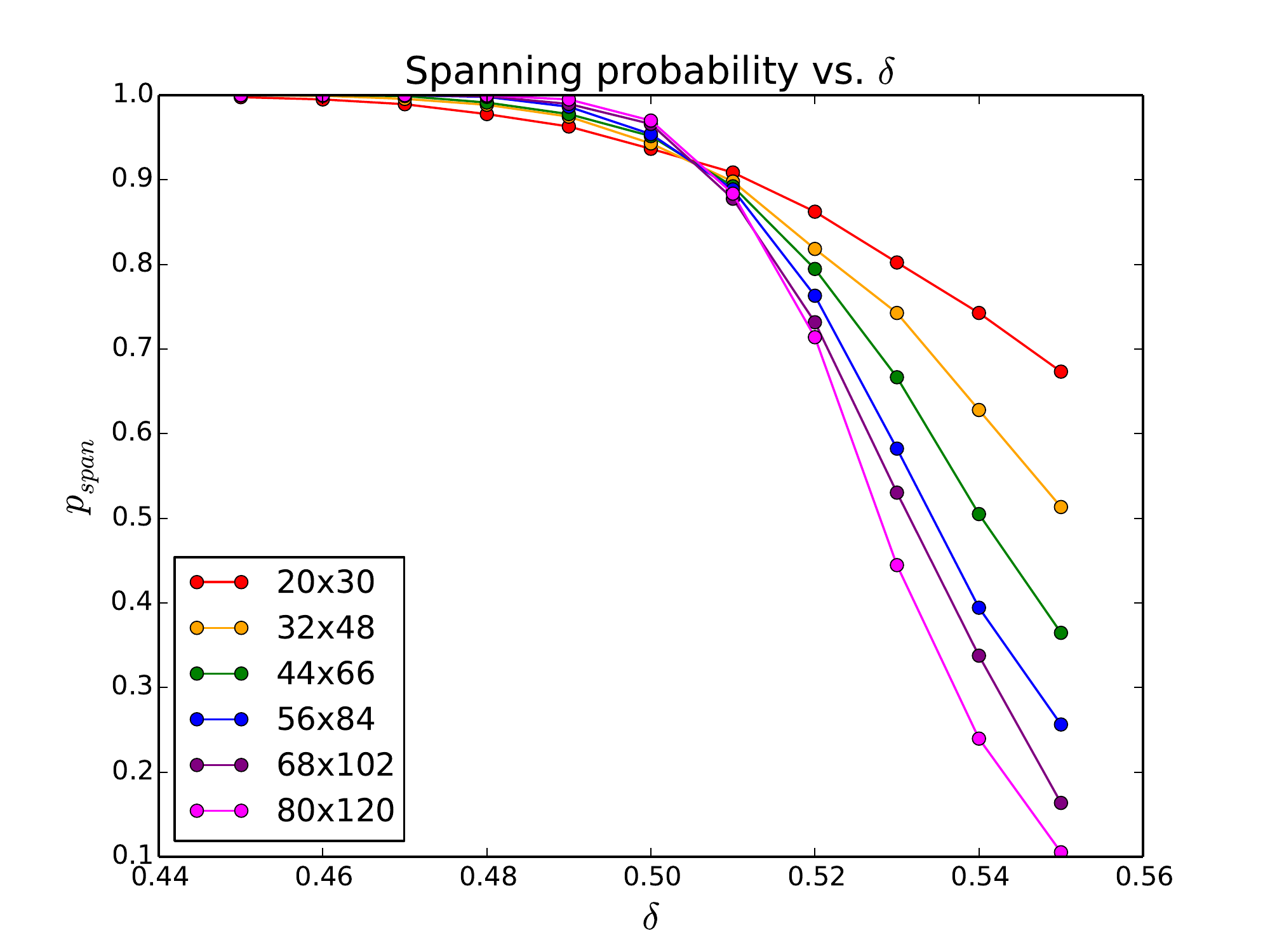}
    \caption{The probability of reducing the ground state of $\aklthamd$ to a well-connected graph state (one with a crossing path) versus the deformation parameter $\delta$ around the transition point, from Monte Carlo simulations. Each line represents a different lattice dimension $L_1\times L_2$. There are three spin-3/2 particles per lattice site, so the total number of particles is $3\times L_1\times L_2$. We observe that below $\delta=0.50\pm0.01$ this probability tends to 1 as the number of particle increases, while above this value the probability tends to 0. We conclude that, below this critical value, the model has a ground state that is universal for MBQC.}
    \label{f:probability_crossing}
\end{figure}

On the other hand, the second condition is only satisfied for values of $\delta$ below some critical value $\delta_c$. The probability $p_{\rm span}$ of the resulting graph state having a crossing path as a function of $\delta$ is plotted in Fig. \ref{f:probability_crossing}. We found a critical value of $\delta_c=0.50\pm0.01$, below which the probability $p_{\rm span}$ increases to 1 as the number of particles increases (i.e.~satisfying the second condition), and above which $p_{\rm span}$ decreases to 0 (violating the second condition). 

As both C1 and C2 are satisfied for $\delta<\delta_c$, the ground ground states of $\aklthamd$ in this region are universal for MBQC. The sudden appearance of a macroscopic graph state at $\delta=\delta_c$ is similar to a percolation phase transition. Given that the existence of a macroscopic graph state implies universality for MBQC, we call the region $\delta<\delta_c$ a computational phase. Whether this phase is characterised by some other physical property other than universality for MBQC is not clear: correlation functions (used to identify the phase boundary in Ref. \cite{niggemann_quantum_1997}) are exponentially decaying for all $\delta\in(0,1]$. One might hope to prove the existence of a quantum phase transition at $\delta=\delta_c$ by showing that the gap (often used as the defining property of a quantum phase) closes at this point, however there does not appear to be a straightforward way to do this. We leave this question open to future investigation.

To summarise, we have shown how the AKLT model on a star lattice, which has a ground state not expected to be universal for MBQC, can be brought into a computational phase where its ground state is universal for MBQC.  In the following section we will show how this approach may be generalised to drive a variety of frustration-free models into computational phases. 

\subsection{General approach for driving frustration-free models into computational phases}
In this section we will outline the general approach for obtaining computational phases of frustration-free spin models. Many features of the star-lattice AKLT model example, described above, can be carried over to the general case. 

As described in Sec. \ref{s:deformationdefs}, our starting model must be frustration-free, and there must exist a set of local rank-2 projectors $(P_j)_{j=1}^N$ that converts its ground state to a graph state $\ket{G}$. We can then define a one-parameter family of models, which we call $\tilde{H}(\delta):\delta\in(0,1]$, which has a ground state that tends towards $\ket{G}$ as $\delta\rightarrow0$. 

While $\ket{G}$ itself may be universal, this alone does not guarantee that the ground state of $\tilde{H}(\delta)$ is universal for any non-zero $\delta$. To show that the ground state of $\tilde{H}(\delta)$ is universal for some non-zero $\delta$, we require a way of dealing with errors that arise from the fact that the ground state is an imperfect version $\ket{G}$, rather than $\ket{G}$ itself. 

We showed that this is possible for the star-lattice AKLT model, described above, because there exists a reduction procedure, defined in Sec. \ref{s:reducing_groundstates} which, for $\delta$ below some critical value $\delta_c$, yields graph states that are universal for MBQC. 

We generalise this reduction procedure as follows. Given the set of projectors $(P_j)_{j=1}^N$ satisfying $\bigotimes_{j=1}^N P_j \groundstate=\ket{G}$, we measure each particle with the two-outcome, site-dependent, projective measurement defined at site $j$ by the two operators $\{P_j, \bar{P}_j:=I-P_j\}$. Note that unlike the AKLT measurement defined in Eq. $\eqref{e:deformed_measurement}$, this measurement is $\delta$-independent and is projective, so that, in principle, entangling to an ancilla is not necessary when measuring individual particles in the resource. However we did not use the general measurement in the AKLT example because the probability distribution for measurement outcomes is not as simple, and there is a larger overhead involved in dealing with unwanted outcomes.

If the $P_j$ outcome is obtained for every $j$, the ground state will be transformed into the graph state $\ket{G}$, however this has an exponentially small probability of success. We will almost always obtain unwanted $\bar{P}$ outcomes, which we will call defects. Let $\sigma$ be a particular outcome configuration, i.e.~a boolean string $\sigma=\sigma_1\sigma_2\dots\sigma_N$ where $\sigma_j=1$ if $P_j$ is obtained at site $j$ and $\sigma_j=0$ if $\bar{P}_j$ is obtained at site $j$. 

Let $p_1(\sigma)$ be the probability of getting a particular outcome configuration $\sigma$ at $\delta=1$. Unlike Eq. \eqref{e:distribution} in the AKLT case, in general it may be difficult to find a simple expression for $p_1(\sigma)$. Nevertheless, given such a $p_1(\sigma)$, it is easy to see how this probability distribution varies with $\delta$. Using the form of the ground state in Eq. \eqref{e:invertible_gs} and the fact that $P_j D_j(\delta)\iv=(1/\delta)P_j$ and $\bar{P}_j D_j(\delta)\iv=\bar{P}_j$, we see that the probability $p_\delta(\sigma)$ that $\sigma$ will be obtained for $\delta\in(0,1]$ is simply given by
\begin{equation}
    p_\delta(\sigma)=\fr{\mathcal{Z}(\delta)}\delta^{-2N_m(\sigma)}p_1(\sigma)\,,
\end{equation}
where $N_m(\sigma)=\sum_{j=1}^N \sigma_j$ is the number of $P$ outcomes obtained and $\mathcal{Z}(\delta)=\sum_\sigma \delta^{-2 N_m(\sigma)}p_1(\sigma)$ is a normalisation factor. The closer $\delta$ is to zero, the larger the term $\delta^{-2}$ will be, and the more probable $P$ outcomes will be, compared with $\bar{P}$. We can reduce the probability of $\bar{P}$ outcomes arbitrarily, by reducing $\delta$, however the probability of obtaining exactly zero $\bar{P}$ outcomes is exponentially small in the number of particles for any fixed $\delta\in(0,1]$. 

Fortunately, a sufficiently small density of defects can be dealt with. Note that this is assuming that the ground state is a PEPS state, such that a defect is effectively isolated at a site, although the argument may hold for more general ground states. Using graph-state update rules, we simply perform a $Z$-basis measurement on particles surrounding a defect to disentangle the defect (as well as the measured particles) from the graph. The resulting state, after defect removal, will be a graph state similar to $\ket{G}$ but with some qubits removed. The fact that we can make the density of defects arbitrarily small by setting $\delta$ sufficiently small and that MBQC with graph states is robust against losing qubits in the preparation stage \cite{browne_phase_2008} implies that, provided $\ket{G}$ is a universal resource, there exists some $\delta>0$ below which the ground states of $\tilde{H}(\delta)$ are universal for MBQC. 

We remark that this general reduction procedure is not necessarily the optimal one for a given spin model. For instance, the star-lattice AKLT model reduction involved a three-outcome measurement \eqref{e:deformed_measurement}, rather than the two-outcome projective measurement used in the general procedure. Obtaining an undesired outcome in the AKLT procedure (one where the outcome is not the same as the deformation axis of the site) modifies the graph in a simple way: usually only a single node and some of its edges are lost. In the general procedure, if an undesired outcome is obtained at a site, not only is that node lost, but all surrounding nodes and their edges must be removed as well. Because of the additional overhead involved in defect removal, the general procedure would require a smaller value of $\delta$ for the resulting graph states to be universal for MBQC. 

To summarise, in this section we defined a reduction procedure, which for sufficiently small $\delta$ converts the ground states of $\tilde{H}(\delta)$ into graph states similar to $\ket{G}$, but with a small density of defects. While this reduction is not necessarily the most efficient one, we can be sure that, provided $\delta$ is small enough, the resulting graph state will be universal for MBQC.

\section{Spectral properties}
\label{s:spectral_properties}
We have seen that, for a member $H$ of particular class of frustration-free Hamiltonians, we can define a one-parameter family of Hamiltonians $\tilde{H}(\delta)$ such that, for sufficiently small $\delta$, the ground state is a universal resource for MBQC. The model was defined such that the ground state of $\tilde{H}(\delta)$ tends to a graph state as $\delta\rightarrow0$, allowing us to use reduction techniques to prove universality. 

An essential consideration for MBQC is whether the resource state can be efficiently prepared, and that it has some robustness against errors. Understanding these properties requires information about excited states in the model, not only the ground state. In this section we will investigate how features of the excited states and the spectrum vary as a Hamiltonian is deformed as described in Sec. \ref{s:deformationdefs}.  

An important property of a local Hamiltonian is its gap. We say that a model is gapped (or has a gap) if the difference between the lowest and second lowest eigenvalues of the Hamiltonian as a function of the number of particles $N$ is lower bounded by some $N$-independent constant. More precisely, if $\Delta_N$ is the difference between the ground state energy and the energy of the first excited state for an $N$ particle Hamiltonian, we define the gap of the model (an $N$-independent quantity) to be $\Delta:=\inf \{\Delta_N\}$, where the infimum is taken over all $N$ for which the Hamiltonian is well defined. 

The gap is a crucial consideration for the efficient preparation and robustness of a resource for MBQC. If the ground state of the undeformed $\delta=1$ Hamiltonian can be obtained efficiently, then as long as $H(\delta)$ is gapped for $\delta\in[\delta_{\rm{min}},1]$, the ground state of $H(\delta_{\rm{min}})$ can also be obtained efficiently by adiabatic evolution.
The notion of a gapped path of Hamiltonians is used to define quantum phases \cite{chen_local_2010}. Two Hamiltonians $H_0$ and $H_1$ are said to be in the same phase if there exists a continuous path of gapped Hamiltonians $H_\gamma$ with $0\le\gamma\le1$ connecting them. According to this definition, smoothly varying the Hamiltonian around a phase, without crossing a phase boundary, results in smooth variation of ground state observables. A central question from the perspective of MBQC is whether we can draw any connection between the phase diagram of a model, and the universality of its ground states for MBQC.

In the following, we will make some general observations about the gap properties of the Hamiltonian, before examining a specific example of a simple 1-D chain.

\subsection{Gap vs. fidelity}
\label{s:gapvsfidelity}

Here we will make some general comments about the gap of our one-parameter family of models defined by Hamiltonians $\tilde{H}(\delta)$ in Sec. \ref{s:deformationdefs}. While $\tilde{H}(\delta)$ can be defined with interactions involving arbitrary numbers of particles, for two-body models the gap must decrease to zero as $\delta$ approaches zero. A simple reason for this is that $\hamd$ has a unique ground state for $\delta>0$, while $\lim_{\delta\rightarrow 0}\hamd$ has an exponentially degenerate ground space, implying that the energy of an exponential number of excited states must decrease to zero as $\delta$ approaches zero. 

We remark that the shrinking of the gap of the two-body Hamiltonian with decreasing $\delta$ can be interpreted as a necessary cost of increasing the fidelity of the ground state with a graph state. Van den Nest previously showed the ground state of a two-body $N$-qubit Hamiltonian can only approximate an $N$-qubit graph state with trade-off between the fidelity and gap \cite{van_den_nest_graph_2008}. While this result is not directly applicable here (as we are not dealing with qubit Hamiltonians) we do observe a similar trade-off. 

However, as we saw in Sec. \ref{s:2-Ddeformation}, for a given model we only need to reduce $\delta$ to some $N$-independent constant for the ground state to be universal for MBQC. For instance, the star-lattice AKLT state we only needed to set $\delta\approx0.5$ to obtain ground states universal for MBQC. Thus, at least for the purpose of universal MBQC, we do not need to shrink the gap by much. Nevertheless, reducing $\delta$ towards zero reduces the randomness of the resulting graph state, and thus reduces the overhead for universal quantum computation. 

Precisely how the gap of the Hamiltonian and the ground state fidelity vary with $\delta$ will depend on the starting Hamiltonian, and the interaction graph. In the following section, we will examine a specific simple example of the spin-1 AKLT model, before commenting on general models.

\subsection{The spin-1 AKLT model}
Analytical proofs of spectral properties, such as a gap, for general spin models are very difficult. Thus we will first look at a simple example of a 1-D chain, where analytical results can be proven exactly, before explaining how results may be generalised. 

\subsubsection{Deforming the spin-1 AKLT model}
\label{s:deformingspin1}
Here we will use the spin-1 AKLT model as a simple example to highlight the effect that the deformation defined in Sec. \ref{s:deformationdefs} has on spectral properties. 
The spin-1 AKLT model in one-dimension is given by 
\begin{equation}
    \akltham=\sum_{j=1}^N P^{J=2}_{j,j+1}\,,
\end{equation}
where $P^{J=2}_{j,j+1}$ is the projection onto the symmetric spin-2 subspace of particles $j$ and $j+1$. We will impose periodic boundary conditions by defining $P^{J=2}_{N,N+1}:=P^{J=2}_{N,1}$. This model is frustration-free and exactly solvable with a unique ground state which we call the spin-1 AKLT state $\akltstate$. This ground state has a simple MPS description \cite{affleck_valence_1988, perez-garcia_matrix_2006}. 

Assuming, for simplicity, that the length of the chain is even, we define a sequence of rank-2 projectors $\left(P_j\right)_{j=1}^N$ as $P_j=S_x^2$ for $j$ odd and $P_j=S_z^2$ for $j$ even, where $S_x$ and $S_z$ are the usual spin-1 operators. These projectors transform $\akltstate$ to $\ket{G}$, a graph state on a ring of length $N$, via 
\begin{equation}
    \ket{G}=\left[\bigotimes_{j=1}^N P_j\right]\akltstate\,,
\end{equation}
where the $j$-th graph state qubit resides on the two-dimensional image of $P_j$. The logical basis $\ket{0}_L,\ket{1}_L$ is $\ket{1}_z, \ket{-1}_z$ on even sites and $\ket{1}_x, \ket{-1}_x$ on odd sites, where $\ket{1}_b, \ket{-1}_b$ are $+1/{-}1$ eigenstates of $S_b$ for $b\in\{x,z\}$. One simple way of showing this is using the tabular form of an MPS described in Ref. \cite{chen_quantum_2010}. Following Sec. \ref{s:deformationdefs}, we can then define a one-parameter family of frustration-free Hamiltonians by $\aklthamd=\sum_{j=1}^N \tilde{h}_{j,j+1}(\delta)$ where 
\begin{equation}
    \tilde{h}_{j,j+1}(\delta) := \mathcal{Q}(\left[D_j(\delta) \otimes D_{j+1}(\delta)\right] P^{J=2}_{j,j+1} \left[D_j(\delta)\otimes D_{j+1}(\delta)\right])\,,
\end{equation}
and where $D_j(\delta):=\delta P_j + (I-P_j)$ for all $j$. The unique ground state $\akltstated$ of $\aklthamd$, has the property that $\ket{\psi_0(1)}=\akltstate$ and $\akltstated\rightarrow\ket{G}$ as $\delta\rightarrow0$. 

We remark that a similar type of deformation to the AKLT model, which preserves frustration freeness and $\mathbb{Z}_2 \times\mathbb{Z}_2$ symmetry, has been studied previously \cite{klumper_matrix_1993}. Correlations, elementary excitations and the entanglement spectra of these models have been studied \cite{bartel_excitations_2003, santos_entanglement_2012}. In contrast to our model, these previous studies consider a deformation that is directed along the $z$-axis on every site (i.e.~$P_j=S_z^2$ for all $j$), rather than along the $z$-axis on even sites and the $x$-axis on odd sites. The ground state for their model in the limit of small $\delta$ is an antiferromagnetic GHZ state, rather than a graph state. 

In the following, we will compare the spectral properties and elementary excitations of our model with this anisotropic antiferromagnet. 

\subsubsection{Fidelity of the deformed ground state}
Before discussing spectral properties of the deformed spin-1 AKLT model, we will relate the $\delta$ parameter to the fidelity of the ground state with a graph state $F=|\langle G \akltstated|^2$. Assuming that $\akltstated$ is normalised, we can use the fact that $\bra{G}=\bra{G}\bigotimes_j P_j$ and that $\bigotimes_j P_j \akltstated \propto \ket{G}$ to see that $F$ represents the probability of obtaining $P_j$ instead of $I-P_j$ on every site when measuring each particle with projective measurement $\{P_j,I-P_j\}$. Therefore $F$ decays exponentially with $N$. 
As discussed previously, and in Ref. \cite{browne_phase_2008}, a graph state can tolerate a non-zero density of defects without destroying its universality for MBQC. Therefore, while $F$ represents the probability of obtaining exactly zero errors, a more useful property in the context of MBQC is $F^{1/N}$. This is an intensive quantity corresponding to the probability that the projective measurement of a given particle $j$ will obtain $P_j$ (assuming that this probability is site-independent). The probability of obtaining $I-P_j$ on a given site $j$ is $\epsilon:=1-F^{1/N}$. Using the MPS form of the ground state, we find the error probability per particle in terms of $\delta$ is given by  
\begin{equation}
\epsilon=1-\frac{2}{\delta^2+2}= \fr{2}\delta^2+O(\delta^4)\,.  
\end{equation}

\subsubsection{Spectral properties of the deformed model}
\label{s:spectral_deformed}
We will now analyse how spectral properties of the model $\aklthamd$, defined above, vary as $\delta$ varies. In particular, we will study the gap and the elementary excitations as $\delta$ is reduced to zero. 

For all $\delta>0$, the two-body interaction $\tilde{h}(\delta)_{j,j+1}$ is a rank-five projection operator. As $\delta\rightarrow0$, the two-body interaction term tends to the projection
\begin{equation}
    \lim_{\delta\rightarrow 0}\tilde{h}_{j,j+1}(\delta)=I-P_j P_{j+1} \,.
\end{equation}
The ground space of this interaction is simply the four-dimensional image of $P_j P_{j+1}$, i.e.~a space of two uncoupled qubits. Therefore, the ground space of the limiting Hamiltonian $\lim_{\delta\rightarrow0}\aklthamd$ has a $2^N$-fold degeneracy and cannot be used as a resource for MBQC. 

\begin{figure*}
    \mbox{\subfigure{\includegraphics[width=0.5\textwidth]{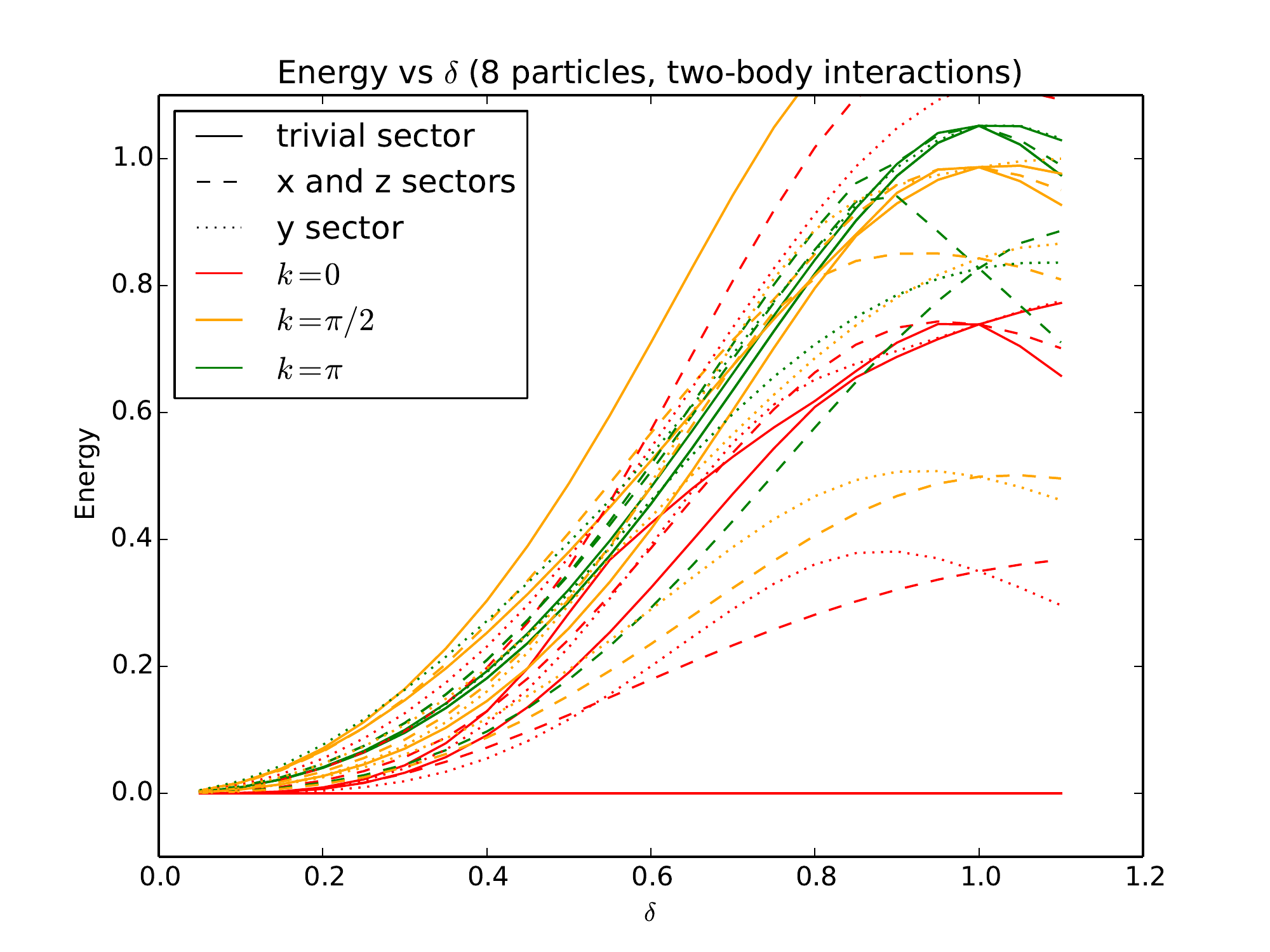}}\subfigure{\includegraphics[width=0.5\textwidth]{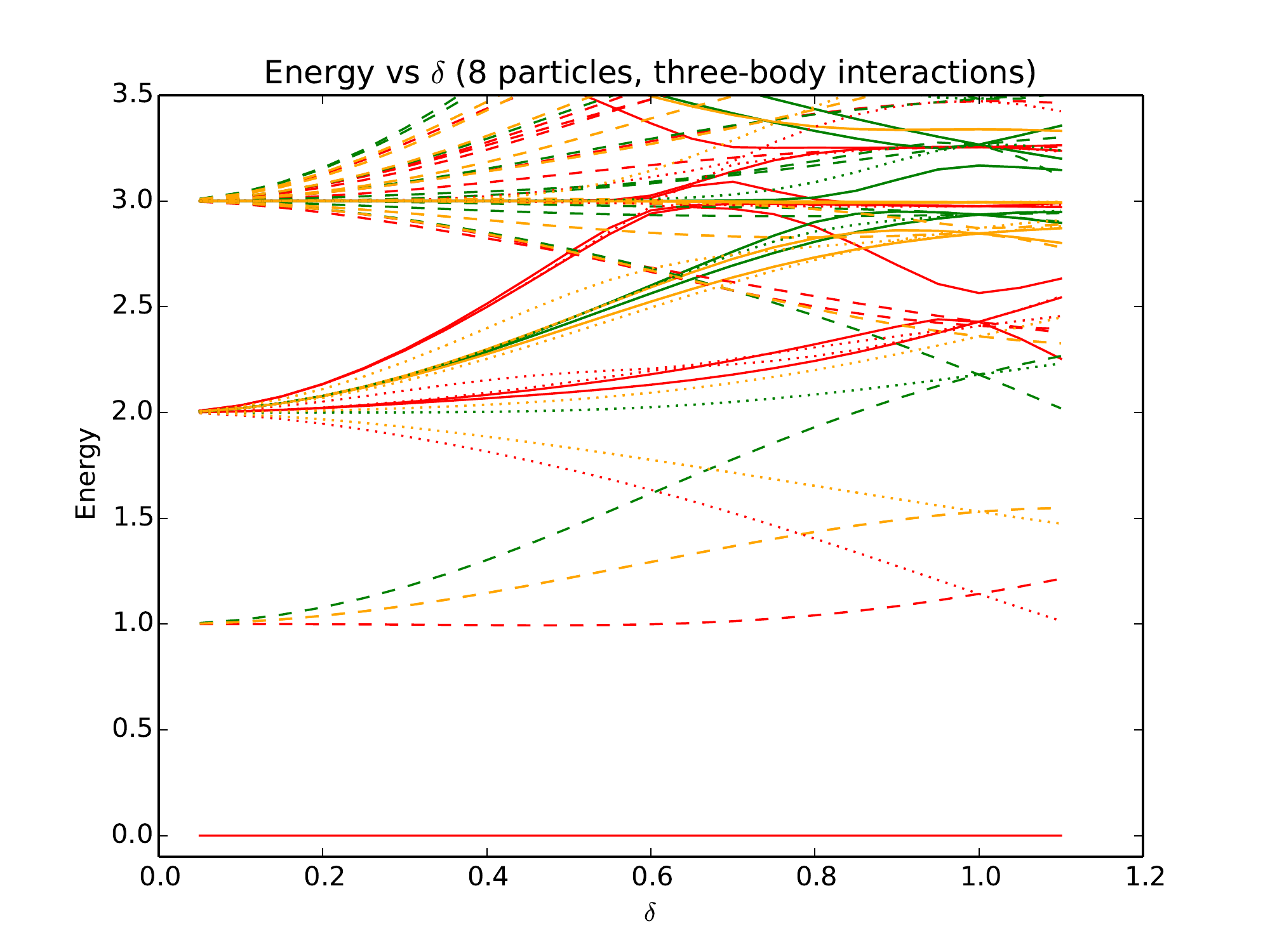}}} 
    \caption{Energy spectrum vs. the deformation parameter $\delta$ for the two-body Hamiltonian $\aklthamd$ (left) and the three-body Hamiltonian $\aklthamthreed$ (right) on a ring of 8 particles.  The ground state for any fixed $\delta$ is identical for both Hamiltonians. Two-site translational symmetry and $\mathbb{Z}_2\times\mathbb{Z}_2$ symmetry have been imposed. The colour of each line indicates the momentum of the eigenstate, where red, orange and green indicate $0$, $\pi/2$ and $\pi$ momentum respectively. The $k=3\pi/2$ levels (not pictured) would be identical to the $\pi/2$ levels due to reflection symmetry. Note that the system has two-site, rather than single-site translational invariance. Thus, $k=\pi$ momentum states under single-site translational invariance (e.g. crackions \cite{knabe_energy_1988}) actually appear to have $k=0$ under two-site translational invariance. For clarity, only the lowest three eigenvalues of each symmetry sector have been plotted in the two-body case. We observe that an exponential number of excited levels approach zero energy in the two-body case. In contrast, as $\delta$ approaches zero, the three-body model approaches the cluster model, which has commuting terms with a unit gap and flat dispersion (i.e.~the same energy for different $k$).}
    \label{f:spectra}
\end{figure*}
To study the spectrum of the model as a function $\delta$, we performed an analysis using MPS methods and exact diagonalisation with up to 14 particles. We have plotted the spectrum of $\aklthamd$ as a function of $\delta$ for a chain of 8 particles in Fig. \ref{f:spectra}.  The spectrum of the 8 particle Hamiltonian is shown rather than the largest calculated 14 particle Hamiltonian as fewer energy levels makes spectral properties more clearly visible. The model is gapped in the thermodynamic limit for any $\delta>0$. We analytically calculated a lower bound for the gap of $O(\delta^4)$ using MPS methods \cite{fannes_finitely_1992, perez-garcia_matrix_2006} and this agrees with our numerics. We can combine this with the expression for the error density $\epsilon=\fr{2}\delta^2+O(\delta^4)$ to relate the error probability per particle to the gap $\epsilon=O(\Delta^{1/2})$. 

We will briefly comment on the elementary excitations of this 1-D model. The elementary excitations of the spin-1 AKLT model are called `crackions' and have been studied numerically and analytically \cite{knabe_energy_1988, arovas_extended_1988, fath_solitonic_1993}. They have also been studied for the AKLT model with a single-axis anisotropy \cite{bartel_excitations_2003} as described above. Crackions can be well approximated by replacing a singlet bond with a triplet state in the valence-bond solid description of the AKLT state, then constructing a momentum eigenstate using all positions of the replaced bond. The crackions have total spin-1, and in the limit of large $N$ have a dispersion of $E(k)=(25+15\cos(k))/27$, with lowest energy at momentum $k=\pi$ \cite{fath_solitonic_1993}.

Using our exact diagonalisation data, we observed the behaviour of the lowest energy $k=\pi$ crackions as $\delta$ was varied. We split energy eigenstates into different $\mathbb{Z}_2\times \mathbb{Z}_2$ symmetry sectors labelled $1,x,y,z$, where $x,y,z$ are -1 eigenstates of global $x,y,z$ flips respectively, and $+1$ eigenstates are unaffected by any flip. The model at $\delta=1$ is fully rotationally symmetric, and the crackions are three-fold degenerate. Reducing $\delta$ from 1 creates anisotropy, causing the $y$ crackion to split from the two-fold degenerate $x$ and $z$ crackions. The energy of each of the $x$, $y$, and $z$ crackions with $k=\pi$ appears to vary as $O(\delta^4)$, and they remain the lowest energy excitations for small $\delta$. The crackions defined for $k\ne\pi$ appear to vary as $O(\delta^2)$. Clearly the model has a non-trivial dispersion relation for small $\delta$. That is, although our model possesses a ground state that approaches a graph state as $\delta\rightarrow 0$, the spectral properties of this model are very different to the cluster model (given by a sum over stabilizers).

To summarise, in this section we highlighted how spectral properties of the one-dimensional AKLT model vary as it is deformed. We showed that the model has a non-trivial dispersion relation for small $\delta$ and the gap shrinks as the fidelity of the ground state with a graph state increases. 

\subsubsection{Two-body vs. three-body interactions}
\label{s:two_vs_three}
In the previous section, we studied the spectral properties of the spin-1 two-body Hamiltonian $\aklthamd$, which has a 1-D graph state as an approximate ground state for small $\delta$. Note that the Hamiltonian deformation defined in Sec. \ref{s:deformationdefs} is not restricted to two-body Hamiltonians. Here we will highlight differences between $\aklthamd$ and higher-body parent Hamiltonians with the same ground state. We observe that the gap of a three-body Hamiltonian with the same ground state does not shrink when the ground state approaches a graph state. This three-body Hamiltonian also has a trivial dispersion relation for small $\delta$. We will see that these properties are inherited from the three-body cluster Hamiltonian on a line.

Let $\aklthamthree=\sum_j h_{j-1,j,j+1}$ be a three-body parent Hamiltonian for $\akltstate$, where each interaction term $h_{j-1,j,j+1}$ is defined as the projection onto the orthogonal complement of the kernel of sums of pairs of neighbouring AKLT interaction terms, i.e.~${h}_{j-1,j,j+1}=\mathcal{Q}(P_{j-1,j}+P_{j,j+1})$. Periodic boundary conditions are imposed by setting $h_{0,1,2}:=h_{N,1,2}$ and $h_{N-1,N,N+1}:=h_{N-1,N,1}$. The resulting Hamiltonian is equivalent to the usual MPS definition of a parent Hamiltonian, where $h_{j-1,j,j+1}$ projects onto the orthogonal complement of the kernel of the reduced density matrix of $\akltstate$ on particles $j-1$, $j$ and $j+1$. Given that $\aklthamthree$ is frustration-free with $\akltstate$ as its unique ground state, we can apply the deformation defined in Sec. \ref{s:deformationdefs} to obtain a three-body Hamiltonian $\aklthamthreed=\sum_{j=1}^N \tilde{h}_{j-1,j,j+1}(\delta)$ that has $\akltstated$ as its unique ground state.

The three-body Hamiltonian $\aklthamthreed$ and the two-body Hamiltonian $\aklthamd$ have the same unique ground state for $\delta>0$. Furthermore, proving that $\aklthamthreed$ is gapped is equivalent to proving that $\aklthamd$ is gapped because 
\begin{align}
    \lambda_{\rm min}(\delta)\tilde{h}_{j-1,j,j+1}(\delta)&\le\tilde{h}_{j-1,j}(\delta)+\tilde{h}_{j,j+1}(\delta)\,,\notag\\
    &\le \lambda_{\rm max}(\delta)\tilde{h}_{j-1,j,j+1}(\delta)\,,
\end{align}
where $\lambda_{\rm min}(\delta)$ and $\lambda_{\rm max}(\delta)$ are, respectively, the smallest and largest non-zero eigenvalues of $\tilde{h}_{j-1,j}(\delta)+\tilde{h}_{j,j+1}(\delta)$. These eigenvalues are independent of $N$ and $j$. More precisely, if $\aklthamthreed$ has gap $\Delta^{(3)}(\delta)$ then $\aklthamd$ has a gap $\Delta^{(2)}(\delta)$ satisfying $\Delta^{(2)}(\delta)\ge\lambda_{\rm min}(\delta)\Delta^{(3)}(\delta)/2$. 

We have plotted the spectrum of $\aklthamthreed$ as a function of $\delta$ next to the spectrum of $\aklthamd$ in Fig. \ref{f:spectra}. We see that the spectrum of $\aklthamthreed$ has a discrete set of evenly spaced energy levels for small $\delta$. As $\delta$ tends to zero, the ground state tends to a graph state. For all $\delta>0$, the interaction term $\tilde{h}_{j-1,j,j+1}(\delta)$ is a projection operator with a four-dimensional kernel equal to the support of the reduced density operator of $\groundstated$ on particles $j-1$, $j$, and $j+1$. Varying $\delta$ can simply be viewed as rotating the kernel of $\tilde{h}_{j-1,j,j+1}(\delta)$. Consider the interaction term obtained by taking the limit $\lim_{\delta\rightarrow 0}\tilde{h}_{j-1,j,j+1}(\delta)$. By continuity, this is a projection operator with a four-dimensional kernel, that contains the graph state $\ket{G}$ in its ground space. The only operator with this property is the three-body parent Hamiltonian of the graph state, given by the three-body stabilizer Hamiltonian 
\begin{equation}
    \lim_{\delta\rightarrow 0}\tilde{h}_{j-1,j,j+1}(\delta)=I-\fr{2}(P_{j-1}P_jP_{j+1}+Z_{j-1}X_j Z_{j+1}) \,,
\end{equation}
where the Pauli operators $X_j$ and $Z_j$ are supported on the logical subspace of $j$ (which we specified in Sec. \ref{s:deformingspin1}). These terms commute and hence the limiting Hamiltonian $\lim_{\delta\rightarrow0}\aklthamthreed$ with $N$ fixed is trivially diagonalisable with unit gap. Furthermore, this Hamiltonian has $\ket{G}$ as its unique ground state. 

Contrast this behaviour to that of the two-body Hamiltonian $\aklthamd$, which has a gap that shrinks to zero as $\delta\rightarrow0$. Recall that $\Delta^{(2)}(\delta)\ge\lambda_{\rm min}(\delta)\Delta^{(3)}(\delta)/2$. Despite the fact that $\Delta^{(3)}\rightarrow1$ as $\delta\rightarrow0$, the gap of the two-body Hamiltonian shrinks because $\lambda_{\rm min}(\delta)=O(\delta^4)$. In other words, the gap of the two-body Hamiltonian shrinks because the smallest non-zero eigenvalue $\lambda_{\rm min}(\delta)$ of the sum of two neighbouring interaction terms $\tilde{h}_{j,j+1}(\delta)+\tilde{h}_{j+1,j+2}(\delta)$ shrinks to zero as $\delta\rightarrow0$. 
In Fig. \ref{f:spectra} we see that the three-body Hamiltonian $\aklthamthreed$, also has crackionic excitations with non-trivial dispersion at $\delta=1$. However, as $\delta$ tends to 0, the dispersion relation of this model becomes completely flat, due to the fact the terms in the limiting Hamiltonian commute. In contrast, as mentioned in the previous section, the dispersion of the two-body Hamiltonian for small $\delta$ is more complicated (it is certainly not flat). 

In this section we showed that there are many simple properties that the three-body parent Hamiltonian $\aklthamthreed$ has that the two-body Hamiltonian $\aklthamd$ does not have, despite both having the same ground state. In particular, the three-body Hamiltonian after taking the limit as $\delta\rightarrow0$ has the graph state as a unique ground state and a unit gap, while the corresponding two-body Hamiltonian has a gap that shrinks as $\delta$ is reduced, and has a degenerate ground space in the limit as $\delta\rightarrow0$. 

\subsection{Generalising spectral properties}
In the previous sections we studied the spectral properties of a simple spin-1 chain that when deformed in the way described in Sec. \ref{s:deformationdefs}, has a linear graph state as an approximate ground state. Here we will discuss whether any of these results can be generalised to systems that have two and higher-dimensional graph states as approximate ground states. 

We explained in Sec. \ref{s:gapvsfidelity} that the gap of any two-body Hamiltonian $\tilde{H}(\delta)$ constructed as in Sec. \ref{s:deformingspin1} shrinks to zero as $\delta$ is reduced to zero and the fidelity of the ground state with any graph state approaches 1. In the particular case of the spin-1 AKLT model, described in the previous sections, we saw that the error density is related to the gap by $\epsilon=O(\Delta^{1/2})$. Unfortunately, describing the trade-off between gap and fidelity, indeed even proving the existence of a gap, becomes highly non-trivial in two and higher dimensions. While there do exist computable sufficient conditions for a frustration-free model to be gapped \cite{perez-garcia_peps_2007, knabe_energy_1988}, these are not always easy to check. For instance, it is still unknown whether the honeycomb lattice spin-3/2 AKLT model is gapped (although there is numerical evidence in favour of it being gapped \cite{garcia-saez_spectral_2013-1}). 

Ultimately, we would like to prove that the model is gapped for sufficiently small $\delta$. If the Hamiltonian $\lim_{\delta\rightarrow0}\tilde{H}(\delta)$ has a graph state as a non-degenerate ground state and a gap (as was the case in the three-body deformed AKLT Hamiltonian in Sec \ref{s:two_vs_three}) then $H(\delta)$ will be gapped for small, non-zero $\delta$. This follows from results concerning the robustness of a gap in PEPS \cite{cirac_robustness_2013-2}. Unfortunately, we cannot directly apply this  result to two-body Hamiltonians, due to the fact that for these models the ground space $\lim_{\delta\rightarrow0}\tilde{H}(\delta)$ is exponentially degenerate. It may be possible to lower bound the gap of a two-body Hamiltonian with the gap of a higher-body Hamiltonian with a simpler spectrum, as we could do in the 1-D case, however the precise details of this remain unclear for general graphs. We leave this question open for future investigation. 

\subsubsection{Comparison to perturbation theory approaches}
The spectral properties studied reveal a number of similarities between our approach and approaches that use perturbation theory to obtain graph states as approximate ground states of two-body Hamiltonians \cite{bartlett_simple_2006, griffin_spin_2008, van_den_nest_graph_2008}. These approaches consider a Hamiltonian of the form $H=gH_S+\lambda V$ where $H_S$ is regarded as the unperturbed Hamiltonian and $\lambda V$ is a small perturbation.  The unperturbed Hamiltonian $H_S$ is exponentially degenerate, much like our limiting Hamiltonian $\lim_{\delta\rightarrow0}\aklthamd$. Adding the perturbation splits the degeneracy of the ground space, selecting an approximation to a graph state $\ket{G}$ as a unique ground state. Much like the $\delta$ parameter in our model, reducing $\lambda/g$ towards zero in their model increases the fidelity of the ground state with the target (encoded) graph state while reducing the size of the gap. 

Both in our approach and in perturbation theory approaches, the graph state qubits reside in two-dimensional subspaces of higher dimensional quantum systems. One difference is that, while perturbation theory approaches involve spin-1/2 particles, our approach requires spin-1 and higher particles.

Furthermore, in the frustration-free models considered in this paper, the ground states have simple MPS/PEPS descriptions. As we saw in the star-lattice AKLT model example in Sec. \ref{s:2-Ddeformation}, this property can be used to obtain an exact expression for the statistics of the reduction procedure and the post-reduction states, thereby allowing us to show that large regions in the parameter space of the model have ground states universal for MBQC. In the perturbation theory approach, errors due to the ground state encoding only an imperfect graph state for non-zero $\lambda/g$ are dealt with using quantum error correction techniques. Hence, while both approaches necessarily result in imperfect graph states as ground states, the imperfection is dealt with quite differently.

While ground space properties are relatively easy to compute in our approach, computing properties of the excited spectrum of the model for small $\delta$ is, in general, non-trivial. We saw in Sec. \ref{s:spectral_deformed} that, even in one dimension, the deformed AKLT model has a complicated dispersion relation for small $\delta$. This is in contrast to the flat spectrum of the cluster model. Perturbation theory approaches yield an entire approximate cluster Hamiltonian in the low energy sector of a two-body Hamiltonian, rather than just an approximate ground state. 

To summarise, our approach yields exactly solvable models with simple tensor network ground states yet complex spectra, while perturbation theory approaches approximate an entire target Hamiltonian albeit with errors that have to be analysed using perturbation theory. 
 
\section{Conclusions}
\label{s:conclusions}
In this paper, we constructed frustration-free two-body Hamiltonians that have graph states as approximate ground states, where each graph state qubit resides in a two-dimensional subspace of each physical particle. The approach works by starting with a frustration-free, two-body Hamiltonian with a ground state that can be stochastically converted to a graph state with local operations, then applying a deformation to the Hamiltonian, parametrised by $\delta$, such that in the limit as $\delta\rightarrow0$, the ground state tends to a graph state. The Hamiltonian remains frustration-free and two-body under this deformation. 

We can use this deformation to obtain phases of frustration-free, two-body Hamiltonians which have ground states that are universal for MBQC. We considered the example of the spin-3/2 AKLT model on a star-lattice, on which existing techniques for performing measurement-based quantum computation fail. Applying this deformation to the star-lattice AKLT model, yeilds an apparent sharp transition in computational power at $\delta\approx0.5$, below which the ground states are universal for MBQC. 

Reducing $\delta$ improves computational properties of the ground state, however, for realistic models that have two-body interactions, it also reduces the size of the gap. Hence there is a trade-off between the fidelity of the ground state with target graph state and the gap of the model. While analytically deriving spectral properties of these types of Hamiltonians is difficult in general, we performed a detailed analysis of the 1-D AKLT model, which under deformation yields a 1-D graph state as an approximate ground state. Restricting to two-body Hamiltonians on a ring we found that the trade-off between the fidelity $F$ of the ground state with a graph state and the gap is given by $1-F^{1/N}:=O(\Delta^{1/2})$. For three and higher body Hamiltonians, there is no such trade-off (fidelity can be improved arbitrarily at no cost). 

New tools are required for a detailed analysis of the gap in two and higher dimensions. One related line of future research would be to examine effect of errors, and non-zero temperature on the computational power of the model. It has been shown that graph states in three dimensions can be used for topological quantum computation with high error thresholds \cite{raussendorf_fault-tolerant_2006, raussendorf_topological_2007} and a natural robustness to thermal noise \cite{raussendorf_long-range_2004, fujii_measurement-based_2013}. It would be interesting if such robustness can be carried over to the graph state models described in this paper, which can also be defined in 3-D.

We remark that this approach to constructing two-body Hamiltonians is quite different to previous approaches that use stabilizers or perturbation theory. It would be interesting to see if this type of construction can be used to define two-body frustration-free models for other information processing tasks. 
\section*{Acknowledgements}
We thank Tzu-Chieh Wei for helpful discussions and for verifying numerical results and Gavin Brennen for helpful comments. This research was supported by the ARC via the Centre of Excellence in Engineered Quantum Systems (EQuS), project number CE110001013. 
\appendix

%
\bibliographystyle{aipnum4-1}

\end{document}